\documentclass[a4paper,12pt]{article}

\usepackage{amsmath,amssymb,slashed,bm,amsfonts,amsthm}
\usepackage{mathrsfs}

\usepackage{url}
\usepackage{xcolor}
\usepackage[colorlinks = true,linkcolor = blue,urlcolor  = cyan,citecolor = blue, anchorcolor = blue]{hyperref}
\usepackage[a4paper,top=2.5cm,left=2.5cm,right=2.5cm,bottom=2.5cm]{geometry}
\usepackage[utf8]{inputenc} 
\usepackage[english]{babel}
\usepackage[T1]{fontenc}
\usepackage{lmodern}
\usepackage{verbatim}
\usepackage{microtype}
\usepackage{titlesec}

\newtheoremstyle{def}{0.2cm}{0.2cm}{\sl}
{16pt}
{\bf}
{{\bf :}}
{ }{}
\newtheoremstyle{rem}
{0.2cm}
{}
{\sl}
{} 
{\bf} 
{{\bf :}} 
{ }
{}

\theoremstyle{rem}
\newtheorem*{remark}{Remark}
\newtheorem{conjecture}{Conjecture}
\newtheorem{example}{Example}
\theoremstyle{def}
\newtheorem{definition}{Def$^{\mathbf{n}}$.}
\newcommand{\rem}{\begin{remark}}
\newcommand{\erem}{\end{remark}}
\newcommand{\conj}{\begin{conjecture}}
\newcommand{\econj}{\end{conjecture}}
\newcommand{\dfn}{\begin{definition}}
\newcommand{\edfn}{\end{definition}}
\newcommand{\be}{\begin{eqnarray}}
\newcommand{\ee}{\end{eqnarray}}
\newcommand{\exa}{\begin{example}}
\newcommand{\eexa}{\end{example}}
\newcommand{\bea}{\begin{eqnarray}}
\newcommand{\eea}{\end{eqnarray}}
\newcommand{\ben}{\begin{eqnarray*}}
\newcommand{\een}{\end{eqnarray*}}

\newcommand{\+}{{+\!\!\!+}}

\newcommand{\im}{\mathrm{i}}

\newcommand{\JJ}{\mathbb{J}}
\newcommand{\VV}{\mathbb{V}}
\newcommand{\XX}{\mathbb{X}}
\newcommand{\YY}{\mathbb{Y}}

\newcommand{\tV}{\widetilde V}
\newcommand{\ty}{\tilde y}

\def\IA{\mathbb{A}}
\def\IB{\mathbb{B}}
\def\IC{\mathbb{C}}
\def\ID{\mathbb{D}}

\def\IR{\mathbb{R}}

\def\IO{\mathbb{O}}
\def\IX{\mathbb{X}}
\def\IV{\mathbb{V}}
\def\IY{\mathbb{Y}}

\def\I1{{1}}

\def\MA{\mathcal{A}}
\def\MB{\mathcal{B}}
\def\MC{\mathcal{C}}
\def\MD{\mathcal{D}}

\newcommand{\pp}{{=\!\!\!|}}

\definecolor{cambridgeblue}{rgb}{0.64, 0.76, 0.68}
\definecolor{lapislazuli}{rgb}{0.15, 0.38, 0.61}
\definecolor{awesome}{rgb}{1.0, 0.13, 0.32}
\definecolor{aureolin}{rgb}{0.99, 0.93, 0.0}
\definecolor{almond}{rgb}{0.94, 0.87, 0.8}
\definecolor{antiquewhite}{rgb}{0.98, 0.92, 0.84}

\newcommand{\CB}[1]{{\color{red} CB: #1}}

\newcommand{\ODD}{\mathrm{O}(D,D)}
\newcommand{\Odd}{\mathrm{O}(d,d)}
\newcommand{\ODDR}{\mathrm{O}(D,D;\mathbb{R})}

\newcommand{\OddR}{\mathrm{O}(d,d;\mathbb{R})}
\newcommand{\OddZ}{\mathrm{O}(d,d;\mathbb{Z})}

\numberwithin{equation}{section}

\newlength{\bibitemsep}
\newlength{\bibparskip}
\let\oldthebibliography\thebibliography
\renewcommand\thebibliography[1]{%
\oldthebibliography{#1}%
\setlength{\parskip}{\bibitemsep}%
\setlength{\itemsep}{\bibparskip}%
}

\begin{document}


\thispagestyle{plain}

\begin{center}
\baselineskip=0pt  

{\Large \bf   Doubled space and extended supersymmetry} 

\vskip 2em
{\bf  Chris D. A. Blair$^{a}$, Ondrej Hulik$^{a}$, Alexander Sevrin$^{a}$, Daniel C. Thompson$^{a,b}$}

\vspace{1em}
{\it  
$^a$Theoretische Natuurkunde, Vrije Universiteit Brussel, and the International Solvay Institutes, \\ Pleinlaan 2, B-1050 Brussels, Belgium \\ 
$^b$Department of Physics, Swansea University, \\ Swansea SA2 8PP, United Kingdom \\ \ \\}

\vspace{1em}
{\tt christopher.blair@vub.be, ondrej.hulik@vub.be, alexandre.sevrin@vub.be, d.c.thompson@swansea.ac.uk}


\end{center}

\begin{abstract} 

The doubled formulation of the worldsheet provides a description of string theory in which T-duality is promoted to a manifest symmetry.  Here we extend this approach to $\mathcal{N}=(2,2)$ superspace providing a doubled formulation for bi-Hermitian/generalised K\"ahler target spaces. The theory is described by a single function, a doubled-generalised K\"ahler potential, supplemented with a manifestly $\mathcal{N}=(2,2)$ constraint. Several examples serve to illustrate this construction, including a discussion of the $\mathcal{N}=(2,2)$ description of T-folds. 
\end{abstract}

{
\hypersetup{linkcolor=black}
\tableofcontents
}


\section{Introduction}
 
The interplay between the physics of the string worldsheet and the geometry of target space is a rich and fascinating subject. 
In this paper, we focus on the supersymmetric worldsheet non-linear $\sigma$-model.
While any target manifold, ${\cal M}$, allows for ${\cal N} = (1,1)$ supersymmetry on the worldsheet, requiring greater amounts of supersymmetry enforces stringent conditions on the metric, $G$, and the Kalb-Ramond two-form $B$ defined on ${\cal M}$.

Our interest is in ${\cal N} = (2,2)$ supersymmetry, as originally investigated long ago \cite{Gates:1984nk}. 
Requiring ${\cal N} = (2,2)$ worldsheet supersymmetry gives particular conditions on the geometrical data in $G$ and $B$.
When the two-form $B$ is absent,  the criteria is that the target space be K\"ahler.   
With the inclusion of $B$, the metric is required to be Hermitian with respect to two (appropriately covariantly constant) complex structures and is accordingly said to be {\em bi-Hermitian}.   

A seminal result of Gualtieri \cite{Gualtieri:2003dx} is that bi-Hermitian geometry can be elegantly reinterpreted as a {\em Generalised K\"ahler Geometry} (GKG) structure on the generalised tangent bundle $E \simeq T{\cal M } \oplus T^* {\cal M}$. With $\dim {\cal M} = D$, the natural pairing between vectors and covectors endows $E$ with an $\ODD$ invariant bilinear pairing, usually denoted by $\eta$.  On $E$, in general and not just for a GKG, the geometric data, $G$ and $B$,  are combined together to form the components of a {\em generalised metric}  denoted ${\cal H}$ or with index raised ${\cal E} = \eta^{-1} {\cal H} $. This generalised metric itself is an $\ODD$ element, and hence obeys ${\cal E}^2 =1$.    In the specialisation to GKG, this data also is encoded in a pair ${\cal J}_{1}$, ${\cal J}_2$ of commuting {\em generalised complex} structures on the generalised tangent bundle.    The generalised metric is identified with the product of the two generalised complex structures, via $ \mathcal{E} =   - \mathcal{J}_1\mathcal{J}_2$.

The natural appearance of $\ODD$ means that the extended bundle $T{\cal M } \oplus T^* {\cal M}$ is relevant more generally in string theory.
It is a natural space on which T-duality can act linearly.  
When $ {\cal M } $ possesses $d$ commuting isometries, such that $T^d \hookrightarrow  {\cal M }  \twoheadrightarrow  {\cal B } $ is a torus fibration over a base manifold ${\cal B } $,  a subgroup $O(d,d; \mathbb{Z})$ of this $ O(D,D) $   becomes the exact T-duality symmetry.\footnote{In supergravity these will both be $\OddR$ and $\ODDR$, but in the full string theory the true T-duality is $\OddZ$. In this work, where the field is left unspecified in a group it should be assumed to be over $\mathbb{R}$.  }

T-duality symmetries are a powerful organisational tool shedding light on the full structure of string theory and its low energy effective descriptions.   Generalised geometry, and the closely related double field theory, have been used to reframe supergravity in a way in which the $O(d,d)$ symmetry becomes a manifest property \cite{Siegel:1993xq,Siegel:1993th,Hull:2009mi}. This is closely related to ideologically similar pursuits in the context of the string {\em worldsheet} theory \cite{Duff:1989tf,Tseytlin:1990va, Hull:2004in,Hull:2006va,Driezen:2016tnz}.
In these worldsheet duality symmetric approaches, the idea is to consider a target space $\mathbb{M}$ that is itself {\em doubled}, i.e. forms a bundle $T^d \times T^d \hookrightarrow \mathbb{M} \twoheadrightarrow  {\cal B } $ in which the torus fibres of which have been doubled (the base space remains the same).  

This goes a step further than generalised geometry as it is the manifold itself (rather than its tangent bundle) that is enlarged. Heuristically, the democratic treatment of momentum and winding modes interchanged by T-duality is accommodated by the inclusion of $d$ additional coordinated directions in this doubled torus target.  An appealing feature of the doubled worldsheet theory is that it may allow for a description of strings in T-folds, a class of non-geometric backgrounds in which locally geometric patches are glued with T-duality transformations.

The purpose of this note is to understand how to combine the requirements of extended supersymmetry with the T-duality symmetric worldsheet.    Given how well adapted concepts of generalised geometry are to both extended supersymmetry and T-duality,  it is of some surprise that the combination of them is not straightforward.
There are, in fact, several technical reasons why this is challenging.   

Firstly, in duality symmetric approaches one typically introduces an extra T-dual boson, $\tilde{x}$, for each boson, $x$, corresponding to a toroidal isometry direction.  In extended supersymmetry however, the real field $x$ is part of a multiplet whose bosonic content contains a second real scalar, $y$, together forming a complex superfield $z = y + i x$.  The $y$ coordinate need not (and generally will not) correspond to some adapted coordinate of an isometry.  We will see that in order to implement the doubled torus fibration specified by fields $\mathbb{X} =( x, \tilde{x})$ we are also compelled to double the base coordinates $\mathbb{Y} = ( y , \tilde{y})$ as well.   Having implemented the doubling one should also have a procedure to constrain the theory so as to not introduce any additional physical degrees of freedom.   On the $\mathbb{X} =( x, \tilde{x})$ this amounts to imposing a worldsheet chirality constraint, $d\mathbb{X} =  \star \mathcal{E}(d\mathbb{X})$,  that encodes the canonical transformation between dual models.  On the $\mathbb{Y}$ we will require a {\em   topological}  constraint $d\mathbb{Y} =   \mathcal{E}(d\mathbb{Y})$. 
In fact, the entire theory can be understood as being defined by a single function--a doubled-generalised-K\"ahler potential $\mathbb{V}(\mathbb{Y})$--in terms of which the topological constraint on the $\mathbb{Y}$ can be integrated to reveal an {\em algebraic} relation that simply undoes the over-doubling so introduced.     

Secondly,  implementing the chirality constraint on the $\mathbb{X}$ at the Lagrangian level is challenging;  doing so in the spirit of Floreanini-Jackiw and Tseytlin requires breaking manifest worldsheet Lorentz invariance which is evidently undesirable from the point of view of the supersymmetry algebra.  The PST approach to chiral bosons does allow one to retain worldsheet Lorentz covariance, at the expense of introducing additional fields, but the extension of this to ${\cal N} =(2,2)$ is very involved  \cite{Driezen:2016tnz,Sevrin:2013nca}.   In this work we will somewhat side step this point and instead, as with Hull's approach to the doubled world sheet \cite{Hull:2004in,Hull:2006va}, view the chirality constraint as an additional ingredient to be imposed off shell by hand.    

The third challenge is there exists a variety of ${\cal N} =(2,2)$ multiplets (chiral, twisted chiral and semi-chiral) and T-duality can change the type of multiplets required to furnish a particular theory.  To simplify the situation we will restrict our attention first to so-called BiLP geometries which are described without semi-chiral multiplets.  Rather appealingly, within our approach once we have appropriately doubled the BiLP geometry we find that we can access T-duality frames in which the conventional description requires semi-chiral multiplets.   

There exist previous studies in the literature making contact between the string worldsheet and generalised complex geometry.
The generalised complex structures of generalised K\"ahler geometry have been shown to appear naturally in the string worldsheet theory in first order or Hamiltonian approaches \cite{Lindstrom:2004iw,Zabzine:2005qf,Bredthauer:2006hf}. Our approach is rather fully covariant on the worldsheet (though see \cite{Lindstrom:2020szt} for an outline of a `covariant Hamiltoninan' alternative).

A doubled superspace formulation exists for certain backgrounds with extended $\mathcal{N}=(4,4)$ supersymmetry. 
This links to the T-folds mentioned earlier.
One of the most studied examples of a T-fold background is the exotic $5_2^2$ brane obtained by T-dualising the NS5 brane on two traverse directions (see \cite{deBoer:2012ma} for a detailed discussion).
This geometry admits extended $\mathcal{N}=(4,4)$ worldsheet supersymmetry and can be studied via an $\mathcal{N}=(2,2)$ description (with a generalised K\"ahler potential) such as in \cite{Kiritsis:1993pb}.
It can alternatively be described via an $\mathcal{N}=(4,4)$ gauged linear sigma model (GLSM).
In \cite{Jensen:2011jna} a doubled $\mathcal{N}=(4,4)$ GLSM was written down capable of describing the duality between the NS5 and Kaluza-Klein monopole, and providing a natural setting with which to discuss worldsheet instanton corrections to the naive Buscher T-duality rules.
Interestingly, here it was already noted that supersymmetry required the doubling of both the fibre and base directions. 
This was later extended to include the $5_2^2$ obtained via a further T-duality in \cite{Kimura:2015qze,Kimura:2018ain}.
The generalised complex structures of this duality chain were investigated very recently \cite{Kimura:2022dma,Kimura:2022jyp}.

The paper is structured as follows:

In section \ref{geometry} we begin with a telegraphic bi-lingual (conventional/generalised) precis of the geometry associated to extended supersymmetry. Here we pay special attention to how generalised geometric considerations lead to a very elegant derivation of the transformations of complex structures under T-duality.  

In section \ref{dwes},   we proceed with the presentation of the $\mathcal{N}=(2,2)$ doubled worldsheet theory and demonstrate that upon reducing to $\mathcal{N}=(1,1)$ superspace it reproduces the existing doubled string Lagrangian and constraints. We elaborate on how $\Odd$ acts in this doubled model, showing in particular that the transformation of complex structures in the doubled space matches that of the usual complex structures of bi-Hermitian geometry.  

The final section, section \ref{examples}, is dedicated to examples displaying how the construction of the doubling works in practice, with some particular applications illustrated.
In subsection \ref{torusexample}, we showcase a simple toroidal model.
In subsection \ref{SU2U1example}, we look at $\mathrm{SU}(2) \times \mathrm{U}(1)$.
In subsection \ref{tfolds}, we discuss the $\mathcal{N}=(2,2)$ realisation of T-folds, and present codimension-1 and codimension-2 examples.
Lastly, in subsection \ref{semi}, we discuss a non-geometric T-duality transformation of a K\"ahler geometry leading to a semi-chiral geometry which we describe without introducing semi-chirals.

\section{The geometry of extended supersymmetry}
\label{geometry}
\subsection{Bi-Hermitian geometry and the ${\cal N}=(2,2)$ worldsheet}
Consider a non-linear sigma-model defined on a target $\{ {\cal M}, G, B\}  $ represented by the Lagrangian 
\begin{equation}
{\cal L} = (G_{IJ}+B_{IJ}) \partial_+ x^I \partial_- x^J\, . 
\end{equation}
Regardless of the choice of metric $G$ and two-form $B$ this theory can be extended to one with ${\cal N}=(1,1)$ supersymmetry by moving to superspace;   replacing $x^I$ with the superfield $X^I(x,\psi, F)$ ($\psi$ a fermion and $F$ an auxiliary boson), and upgrading the derivatives $\partial_\pm$ to super-covariant derivatives $D_\pm $ and integrating:
\begin{equation}
\label{susysigma}
{\cal L} =\int {\rm d}^2 \theta \,  (G_{IJ}+B_{IJ}) D_+ X^I D_- X^J\, . 
\end{equation}
One can then postulate the existence of an  additional supersymmetry that mixes these superfields  
\begin{equation}
\delta X^I = \epsilon_+ (J_+)^I{}_J D_+ X^J   + \epsilon_- (J_-)^I{}_J D_- X^J\, .
\end{equation}
The analysis of \cite{Gates:1984nk}  shows for this transformation to be consistent that $J_\pm$ are complex structures obeying, for all vectors $U,V \in \Gamma( T{\cal M}) $, 
\begin{equation}
\label{eq:biH1}
J_+^2=J_-^2=-1 \,,\qquad [U,V]+J_\pm[J_\pm U,V]+J_\pm[U,J_\pm V]- [J_\pm U,J_\pm V]=0\,.
\end{equation}
To leave the Lagrangian invariant under the additional supersymmetry we further require that 
the metric $G$ is Hermitian with respect to both complex structures 
\begin{equation}
\label{eq:biH2}
G(J_\pm U,J_\pm V)=G(U,V)\,, 
\end{equation}
and that the complex structures define  two-forms $\omega_\pm(U,V)=-G(U,J_\pm V)$ which satisfy,
\begin{equation}
\label{eq:biH3}
\mathrm{d} \omega _\pm (U,V,W) =\mp H(J_\pm U,J_\pm V,J_\pm W)\,,
\end{equation}  
for any three vectors $U,V,W$.
The last condition is equivalent to the requirement that the complex structures are covariantly constant with respect to the Levi-Civita connection $\pm$ torsion $H = dB$. A target space possessing these properties is often called a bi-Hermitian geometry.

To manifest this second supersymmetry off-shell requires the introduction of ${\cal N} = (2,2)$ superspace (see appendix for details).  There are three pertinent superfields and which should be used depends crucially on the properties of $J_\pm$:  chiral fields $z^\alpha, \bar{z}^\alpha$ parametrise  $\mathrm{ker}(J_+ - J_- )$; twisted chiral fields $w^\mu, \bar{w}^\mu$ parametrise   $\mathrm{ker}(J_+ + J_- )$; and semi-chiral fields  \cite{Buscher:1987uw} $(l ,r , \bar{l} ,\bar{r})$  parametrise the remaining directions $\textrm{im}( [J_+ , J_-] G^{-1})$.  Should the target geometry be of the type that only chiral and twisted chiral superfields are required, i.e. $[J_+,J_-]=0$, then it is said to be BiLP (bi-Hermitian local product). This BiLP class, which we will largely focus on in the present work, clearly encompasses K\"ahler geometry as a special case where $J_+ = J_-$. 

In ${\cal N}=(2,2)$ superspace, a single real function known as the generalised K\"ahler potential,    $V(z,w,\bar z, \bar w)$ for a BiLP,  specifies the Lagrangian 
\begin{equation}
{\cal L} = \int {\rm d}^2\theta {\rm d}^2 \hat \theta \, V(z,w,\bar z, \bar w) \,.
\label{SVorig}
\end{equation}
Passing back to ${\cal N}=(1,1)$ superspace (discarding total derivatives) one obtains\footnote{We use the notation $V_{\alpha} \equiv \frac{\partial V}{\partial z^\alpha}$, $V_{\bar \alpha} \equiv \frac{\partial V}{\partial \bar z^\alpha}$, and so on.}
\be
\begin{split}
\mathcal{L} & = 
2 V_{\alpha \bar \beta} ( D_+ z^\alpha D_- \bar z^\beta + D_+ \bar z^{\beta} D_- z^\alpha)
- 2V_{\mu \bar \nu} ( D_+ w^\mu D_- \bar w^\nu + D_+ \bar w^{\nu} D_- w^\mu )
\\ & \qquad
- 2 V_{\alpha \bar \mu} ( D_+ z^\mu D_- \bar w^{\mu} - D_+ \bar w^{\mu} D_- z^{\mu}) 
- 2 V_{\bar\alpha  \mu} ( D_+ \bar z^\mu D_-  w^{\mu} - D_+  w^{\mu} D_- \bar z^{\mu}) \,,
\end{split}
\ee
which shows that the  metric and B-field are extracted as derivatives of this potential according to\footnote{The perhaps unseemly factors of 2 are to match with our later conventions involving real coordinates.}
\begin{equation}
G_{\alpha \bar \beta} = 2 V_{\alpha \bar \beta} \,,\quad
G_{\mu\bar \nu} = - 2 V_{\mu \bar \nu} \,,\quad
B_{\alpha \bar \mu} = - 2V_{\alpha \bar \mu} \,,\quad
B_{\bar \alpha \mu} = - 2V_{\bar \alpha \mu} \,.
\end{equation} 
If in addition semi-chiral multiplets are required similar, albeit significantly more involved, expressions are found for the geometric data in terms of derivatives of the generalised K\"ahler potential.

\subsection{Generalized K\"ahler Geometry}
\label{genkahgeometry}

We now briefly review how the bi-Hermitian geometry invoked by ${\cal N}=(2,2)$ supersymmetry can be reformulated  in terms of  generalized K\"ahler geometry on the  generalized tangent space $E\simeq T{\cal M } \oplus T^\star {\cal M}$ \cite{Gualtieri:2003dx} \cite{Hitchin:2004ut}.  Given two sections of this bundle $\mathbb{U} = (U, \alpha)$ and $\mathbb{V} = (V, \beta)$ we can construct the $\ODD$ invariant paring 
\begin{equation}
\eta( \mathbb{U},\mathbb{V} ) = \iota_U \beta + \iota_V \alpha 
\end{equation}
and the $H$ twisted Courant bracket
\begin{equation}
[ \mathbb{U} ,\mathbb{V}]_H := [U,V]+\iota_U {\rm d}\beta- \iota_V {\rm d} \alpha+ \tfrac{1}{2} {\rm d} \big(\iota_U\beta - \iota_V\alpha \big) + \iota_V\iota_U H\,. 
\end{equation}
Given a metric and two-form on ${\cal M}$ we can also endow $E$ with a generalised metric ${\cal H}$, a representative of the coset $\ODD/ O(D)\times O(D)$ that provides an additional symmetric pairing.   We shall often work in a basis of $E$ for which the inner products are given by 
\begin{equation} 
\eta= \begin{pmatrix}
0& \I1_{D} \\ \I1_{D} & 0
\end{pmatrix}    \, , \quad  
\mathcal{H} = \begin{pmatrix}
G - B G^{-1} B & - B G^{-1} \\
G^{-1} B & G^{-1} 
\end{pmatrix} \,.
\end{equation}
In places it is more natural to view the generalised metric as an endomorphism of $E$ defined by $\mathcal{E}= \eta^{-1} \cdot \mathcal{H} $. As $ \mathcal{H}$  is an $\ODD$ element it follows that $\mathcal{E}^2 =\textrm{id} $ and hence defines projectors 
\be 
{\cal P}_\pm = \frac{1}{2} \left( \textrm{id} \pm {\cal E} \right) \, . 
\ee 
In general ${\cal E}$ can be decomposed as
\begin{eqnarray} 
\mathcal{E}= \left( \begin{array}{cc}\I1_{D}&0\\-B&\I1_{D} \end{array}\right) \left( \begin{array}{cc}0&G^{-1}\\G&0 \end{array}\right) \left( \begin{array}{cc}\I1_{D}&0\\B&\I1_{D} \end{array}\right)\,.\label{prodstr} 
\end{eqnarray} 
A generalised K\"ahler structure on $E$ consists of a pair,   ${\cal J}_1$ and ${\cal J}_2$, of  generalised complex structures such that $  
{\cal J}_1^2 = {\cal J}_2^2=-\textrm{id}$ and $ [{\cal J}_1 , {\cal J}_2  ] =0$ which are Courant-integrable, 
\be 
[  \mathbb{U}, \mathbb{V} ]_H +{\cal J}_{1,2} [ {\cal J}_{1,2}  \mathbb{U}, \mathbb{V} ]_H +{\cal J}_{1,2} [  \mathbb{U},{\cal J}_{1,2}  \mathbb{V} ]_H - [{\cal J}_{1,2} 
\mathbb{U},{\cal J}_{1,2}  \mathbb{V} ]_H =0\, \quad \forall \mathbb{U}, \mathbb{V} \in \Gamma(E) \, .\label{six}
\ee 
Furthermore, a  generalized metric can now be obtained from the product of two generalized complex structures
\be 
{\cal E}&=&- {\cal J}_1 {\cal J}_2   \,.
\ee  
A remarkable result is that the conditions of bi-Hermitian geometry of eqs. \eqref{eq:biH1}-\eqref{eq:biH3} can be equated to the definition of a generalised complex structure.  This is achieved explicitly through the Gualtieri map\footnote{In which ${\cal J}_1$ corresponds to the lower choice in $\pm$. } 
\begin{eqnarray}	
{\cal J}_{1,2}= \frac 1 2\left( \begin{array}{cc}\I1_{D}&0\\-B&\I1_{D} \end{array}\right)
                                    \left( \begin{array}{cc}
                                    J_+ \pm J_- & \omega^{-1}_+ \mp \omega^{-1}_- \\
                                    -(\omega_+ \mp \omega_-) & -(J^t_+ \pm J^t_-)
                                    \end{array}\right)
                                    \left( \begin{array}{cc}\I1_{D}&0\\B&\I1_{D} \end{array}\right) \, . 
\end{eqnarray}

\subsection{Transformation of complex structures under T-duality}
\label{transJT}
A virtue of the generalised K\"ahler perspective is that the   action of $\ODDR$, i.e. the transformations $\mathscr{O}$ that preserve $\eta =\mathscr{O}^t \eta \mathscr{O} $, have a linear action on the tensors:
\be
\mathbb{U} 
\rightarrow \mathscr{O}^{-1}\mathbb{U} 
\,,
\quad
{\cal J}_{1,2} \rightarrow \,\mathscr{O}^{-1} {\cal J}_{1,2} \mathscr{O}\,,
\quad 
{\cal E} \rightarrow \,\mathscr{O}^{-1} {\cal E} \mathscr{O}\,,
\quad 
{\cal H}\rightarrow  \mathscr{O}^{t}{\cal H}\, \mathscr{O}\,.
\label{genc}
\ee
In this way the complicated T-duality transformations -- which involve a subgroup of the full $\ODDR$ acting only in isometry directions -- of geometric quantities, $G,B, J_\pm$, can be made transparent.  Here we illustrate this by rederiving the transformation rules under T-duality of the spacetime complex structure  originally worked out by Hassan \cite{Hassan:1994mq}.

Now let's take the spacetime metric $G = (G_{IJ})$ and choose for this a vielbein  
$e=(e^A{}_I )$ and flat metric $h=(h_{AB})$  such that  
$  
G=e^th e$.   In terms of this we can now construct a natural basis  $\{ {\cal V}_{+A} ,\, {\cal V}_{-A} \}$, $A\in\{1,\cdots,D\}$,  for the generalized tangent bundle $E\simeq T {\cal M}\oplus T {\cal M}^*$,  
\begin{eqnarray} 
{\cal V}_\pm=
\frac{1}{\sqrt{2}} \,\left( \begin{array}{cc}\I1&0\\-B&\I1 \end{array}\right)\left( \begin{array}{c}\pm e^{-1}\\ e^th \end{array}\right)\,, 
\end{eqnarray} 
and its   dual   $\{ \bar{\cal V}_+^A, \,\bar{\cal V}_-^A\}$ , 
\begin{eqnarray} 
\bar{\cal V}_\pm=
\frac{1}{\sqrt{2}} \,\left( \begin{array}{cc}e \,, \pm h^{-1}e^{-t} \end{array}\right)\left( \begin{array}{cc}\I1&0\\B&\I1 \end{array}\right)\,.  
\end{eqnarray} 
This basis is orthonormal in the sense that:
\be
\bar{\cal V}_+^A {\cal V}_{+B} = \delta ^A_B \,,\quad
\bar {\cal V}_-^A {\cal V}_{-B}=- \delta ^A_B\,,
\quad
\bar{\cal V}_+^A {\cal V}_{-B} =0 \,,\quad
\bar{\cal V}_-^A {\cal V}_{+B} =0 \,,
\ee
and enjoys the completeness relation:
\begin{eqnarray} 
\sum_A {\cal V}_{+A} \otimes \bar {\cal V}^A_+ - {\cal V}_{-A}  \otimes  \bar{\cal V}^A_-=\I1_{2D}\,. 
\end{eqnarray} Furthermore we can reconstruct the previously defined projectors via $   
\sum_A {\cal V}_{\pm A} \otimes \bar {\cal V}^A_\pm=  {\cal P}_\pm$.  

As these vielbein inherit  the natural action of $\ODDR$ i.e. $ \mathcal{V} _\pm \rightarrow \tilde{\mathcal{V}}_\pm = \mathscr{O}^{-1} \mathcal{V}_\pm $, $ \bar{\mathcal{V}} _\pm \rightarrow 
\tilde{\bar{\mathcal{V}}}_\pm = \bar{\mathcal{V}}_\pm \mathscr{O} $,   we can use their transformation to deduce the T-duality rules on $e$ to find a T-dual veilbein $\tilde{e}$. In fact, depending on if we consider  $\mathcal{V} _-$ or  $\mathcal{V} _+$  we find two different results for the T-dual veilbein $\tilde{e}_\pm$.   This is neither a contradiction nor a surprise; as  $ \tilde{e}_+^t h \tilde{e}_+ =  \tilde{e}_-^t h \tilde{e}_-  $  we have that    $\tilde{e}_\pm$ are related by a local Lorentz transformation reflecting the fact that on the worldsheet left and right movers transform differently under T-duality.   To make this precise we   
we will need to use some facts about the structure of transformations $\mathscr{O}\in O(D,D;\IR)$.
Representing the matrix $\mathscr{O}$ in a block diagonal form, 
\begin{eqnarray} 
\mathscr{O} =
\begin{pmatrix}
\IA &\IC \\ \IB&\ID 
\end{pmatrix} \,,
\label{defO1} 
\end{eqnarray} 
one verifies that $O(D,D)$ invariance implies
\begin{eqnarray} 
\IB^t\IA+\IA^t\IB=0=\ID^T\IC+\IC^t\ID\,,\qquad \IB^t\IC+\IA^t\ID=\I1_{D}\,,\label{les1} 
\end{eqnarray} 
or equivalently, 
\begin{eqnarray} 
\IA\IC^t+\IC\IA^t=0=\IB\ID^t+\ID\IB^t\,,\qquad \IB\IC^t+\ID\IA^t=\I1_{D}\,.\label{les2} 
\end{eqnarray} 
Then, starting from $\mathcal{V}_+ \rightarrow \mathscr{O}^{-1} \mathcal{V}_+$ one gets,
\begin{eqnarray}
e \rightarrow \tilde e_+ = e\left(\IC^t E^t+\ID^t\right)^{-1}\equiv e\,\IO_+^{-1}\,,\label{otp}
\end{eqnarray} 
and for $E=G+B$ one finds,
\begin{eqnarray}
E \rightarrow \tilde E= (E\IC+\ID)^{-1}(E\IA+\IB)\,.\label{bes1}
\end{eqnarray} 
On the other hand when starting from $\mathcal{V}_-$ one gets,
\begin{eqnarray}
e \rightarrow \tilde e_- = e\left(-\IC^t E +\ID^t\right)^{-1}\equiv e\,\IO_-^{-1}\,,\label{otm}
\end{eqnarray} 
and for $E$ one finds,
\begin{eqnarray}
E \rightarrow \tilde E= (\IA^tE-\IB^t)(-\IC^tE+\ID^t)^{-1}\,.\label{bes2}
\end{eqnarray} 
Using the identities eqs.~(\ref{les1}) and (\ref{les2}) one easily shows that the transformation rules in  eqs.~(\ref{bes1}) and (\ref{bes2}) are the same. 
One additionally has that $\Lambda = e \IO_+^{-1} \IO_- e^{-1}$ is the Lorentz transformation relating $\tilde e_+$ and $\tilde e_-$.   The utility of this is revealed by noting that we can construct $\ODD$ invariant quantities by   contracting generalised tensors with the above generalised vielbeins and in particular we have 
\be
\bar{\mathcal{V}}^A_\pm {\cal J}_1 \mathcal{V}_{\pm B} =\pm  (e J_\pm e^{-1})^A{}_B\, , \quad \bar{\mathcal{V}}^A_\pm {\cal J}_2 \mathcal{V}_{\pm B} =   (e J_\pm e^{-1})^A{}_B\, , \quad \bar{\mathcal{V}}^A_\pm {\cal J}_i \mathcal{V}_{\mp B} =  0\,  .    
\ee 
As these are inert under $\mathrm{O}(D,D)$ transformations, 
using equations \eqref{otp} and \eqref{otm}, we immediately read off the transformation rules for the complex structures $J_\pm$:
\begin{eqnarray} 
J_\pm \rightarrow \tilde J_\pm= \mathbb{O}_\pm J_\pm \mathbb{O}_\pm^{-1}\,. 
\label{jstrsf} 
\end{eqnarray}

\paragraph{$\Odd$ versus $\ODD$}
To describe genuine T-duality transformations in this language, we assume that $T^d \hookrightarrow M \twoheadrightarrow B $ and the vector fields  that generate the torus action should be isometries of $G$ and $H=dB$. In this case a subgroup $O(d,d; \mathbb{Z})$ of this $ O(D,D) $ relates equivalent string theory backgrounds. 
For our purpose we will be slightly more restrictive by assuming the two-form potential  $B$ is also invariant\footnote{In any case, even to speak of this potential means that we are working in patchwise fashion.}   and so too the  $J_\pm$. We  choose adapted coordinates such that all geometric data,  {\em i.e.} $G$, $B$ and $J_\pm$, are independent of the adapted coordinates. We denote these adapted coordinates by $x^i$, $i=1,\dots,d$, and the remaining non-isometric coordinates by $y^a$, $a=1,\dots, D-d$.
We then focus on the subgroup $O(d,d;\IR)\subseteq O(D,D;\IR)$ with $d\leq D$. The explicit embedding of $\mathcal{O}\in O(d,d;\IR)$ in $ O(D,D;\IR)$ is given by eq.~(\ref{defO1}) where, 
\begin{eqnarray} 
&&\IA=\left( \begin{array}{cc}\MA &0\\ 0&\I1_{(D-d)}\end{array}\right)\,,\qquad 
\ID=\left( \begin{array}{cc}\MD&0\\ 0&\I1_{(D-d)}\end{array}\right)\,, \nonumber\\ 
&&\IB=\left( \begin{array}{cc}\MB&0\\ 0&0_{(D-d)}\end{array}\right)\,,\quad 
\IC=\left( \begin{array}{cc}\MC&0\\ 0&0_{(D-d)}\end{array}\right)\,, 
\end{eqnarray} 
such that we can define an element $\mathcal{O}$ of $\OddR$ given by
\begin{eqnarray} 
\mathcal{O}\equiv \left( \begin{array}{cc}\MA&\MC\\ \MB&\MD\end{array}\right)\,.
\end{eqnarray}

\paragraph{Example: Buscher transformations} 

The particular case of a Buscher transformation in a particular direction labelled by $i$ corresponds to:
\be
\mathcal{A} = \mathcal{D} = \I1_{d} - \varrho_i\,,\quad
\mathcal{B}= \mathcal{C} =  \rho_i \,,\quad 
\ee
where $(\varrho_i)_{jk}= \delta _{ij} \delta_{ik}$, $i,\,j,\,k\in\{1, \cdots,d\}$. 
The Buscher transformation in all $d$ isometry directions corresponds is given by the product of all such individual Buscher transformations, and corresponds to:
\be
\mathcal{A} = \mathcal{D} = 0\,,\quad
\mathcal{B}= \mathcal{C} = \I1_{d} \,. 
\ee
In this latter case, writing $E=G+B$ as 
\begin{eqnarray} 
E=\left( \begin{array}{cc}E_{xx}&E_{xy}\\E_{yx}& E_{yy} \end{array}\right)\,, 
\end{eqnarray} 
(where we let $x$ and $y$ schematically denote the isometry and non-isometry indices)
we get from eq.~(\ref{bes1})\,, 
\begin{eqnarray} 
E \rightarrow \tilde E= \left( \begin{array}{cc}E_{xx}^{-1}&E_{xx}^{-1}E_{xy}\\-E_{yx}E_{xx}^{-1}&E_{yy}-E_{yx}E_{xx}^{-1}E_{xy} \end{array}\right)\,, 
\end{eqnarray} 
which are exactly the Buscher rules \cite{Buscher:1987sk,Buscher:1987qj}. 
The complex structures then transform as in eq.~(\ref{jstrsf}) where $\IO_\pm$ are given by, 
\be
\IO_+ = 
\left( \begin{array}{cc}E_{xx}^t&E_{yx}^t\\0& \I1 \end{array}\right)\,,\quad
\IO_-=
\left( \begin{array}{cc}-E_{xx}&-E_{xy}\\0& \I1 \end{array}\right)\,. 
\ee
If $J_+$ and $J_-$ commute, then this property is preserved by Buscher transformations.

\paragraph{Example: bivector transformation of a K\"ahler geometry}
We present a simple example where a T-duality transformation renders commuting complex structures non-commuting.
Let us consider a $D=2d$-dimensional K\"ahler geometry, with $d$ isometries.
The complex structures are equal, $J_+ = J_-$.
Coordinates can be chosen such that metric takes the form $G_{xx} = G_{yy} \equiv g$, $G_{xy} = 0$.
Hence we can write 
\be
J_+ = J_- = \begin{pmatrix} 0 & \I1_{d } \\
-\I1_{d } & 0 \end{pmatrix} \,,\quad
G = \begin{pmatrix} g & 0 \\ 0 & g \end{pmatrix} \,.
\label{KahlerExample}
\ee
There is no $B$-field.

We consider an $\Odd$ transformation generated by
\be
\mathcal{A} = \mathcal{D} = \I1_{d}\,,\quad
\mathcal{B} = 0 \,\quad
\mathcal{C} = \beta \,,
\ee
where $\beta^T = - \beta$ has the interpretation of a bivector.
The transformation matrices for the complex structures are
\be
\mathbb{O}_+ = \begin{pmatrix}
\I1_{d } - \beta g & 0\\
0 & \I1_{d }
\end{pmatrix}
\,,\quad
\mathbb{O}_- = \begin{pmatrix}
\I1_{d } + \beta g & 0 \\
0 & \I1_{d }
\end{pmatrix} \,.
\ee
leading to
\be
J_+' = \begin{pmatrix} 0 & \I1_{d} - \beta g \\ - (\I1_{d}-\beta g)^{-1}&0 \end{pmatrix} \,,\quad
J_-' = \begin{pmatrix} 0 & \I1_{d} + \beta g \\ - (\I1_{d}+\beta g)^{-1} \end{pmatrix} \,.
\label{JSCHPRIME1}
\ee
These do not commute for any non-zero $\beta$. 
Hence the resulting geometry, with metric and $B$-field given by
\be
G'+B' = \begin{pmatrix} 
g'+ b' & 0 \\
0 & g
\end{pmatrix} 
\,,\quad
g' + b' = (g^{-1} + \beta)^{-1} \,,
\label{GBSCHPRIME1}
\ee
will (ordinarily) be described by semi-chiral superfields in an $\mathcal{N}=(2,2)$ description.

We can use coordinate transformations of the non-isometric coordinates $y$ (which are unaffected by the T-duality) to simplify one or other of the two complex structures  \eqref{JSCHPRIME1}.
For instance, if we define new coordinates
\be
y' = y'(y) \,,\quad
\text{such that}\quad
\frac{\partial y'}{\partial y} = 1 -\beta g 
\,,
\ee
then we obtain 
\be
J_+' = \begin{pmatrix} 0 & \I1_{d}  \\ - \I1_{d}&0 \end{pmatrix} \,,\quad
J_-' = \begin{pmatrix} 0 & (\I1_{d} + \beta g)(\I1_{d}-\beta g)^{-1}\\ - (\I1_{d}-\beta g)(\I1_{d}+\beta g)^{-1}&0 \end{pmatrix}\,.
\label{JSCHPRIME2}
\ee
This corresponds to diagonalising $J_+'$ in a complex basis. It also brings the metric and $B$-field into a more symmetric form with 
\be
G'+B' = \begin{pmatrix} 
g'+ b' & 0 \\
0 & g' 
\end{pmatrix} 
\,,\quad
g' + b' = (g^{-1} + \beta)^{-1} \, ,
\label{GBSCHPRIME2}
\ee
in which one notices in particular that the base-base component of $G'+B'$ has been transformed.   In later examples, a democracy between the fibre and base will be restored by adding also $b'$ as a   two-form potential into this base-base component - however such a change will arise as the incorporation of a total derivative on the worldsheet.


\section{Doubled worldsheet for extended supersymmetry}
\label{dwes}

\subsection{Review of \texorpdfstring{$\mathcal{N}=(1,1)$}{(1,1)} sigma model}
In this section we review how one can rewrite a $\mathcal{N}=(1,1)$ sigma model in an ${\rm O}(d,d)$-covariant language. 
The $\mathcal{N}=(1,1)$ non-linear $\sigma$-model Lagrangian is given by \eqref{susysigma}
We may introduce a split of the coordinates $X^I = (y^a, x^i)$, with $i=1,\dots d$, and decompose the metric and two-form as
\be
G_{ab} = g_{ab} + A_a{}^i A_b{}^j g_{ij} \,,\quad
G_{ai} = g_{ij} A_a{}^j \,,\quad
G_{ij} = g_{ij}\,, 
\ee
\be
B_{ab} = b_{ab} + A_{[a}{}^i A_{b] i} + b_{ij} A_a{}^i A_b{}^j \,,\quad
B_{ai} = - A_{ai} - b_{ij} A_a{}^j \,\quad
B_{ij} = b_{ij} \,.
\ee
In practice, we will consider manifolds with isometries such that the $x^i$ are adapted coordinates for the isometries, along which we can T-dualise, and refer to the $y^a$ as `spectators'. 

The Lagrangian \eqref{susysigma} can be written in terms of this split as
\be
\begin{split}
\mathcal{L} 
& = ( g_{ab} + b_{ab} - A_{[a}{}^i A_{b]i} ) D_+ y^a D_-y^b 
+ A_{ai} (\nabla_+ x^i D_- y^a - D_+y^a \nabla_- x^i ) 
\\ & \qquad + (g_{ij} + b_{ij}) \nabla_+ x^i \nabla_- x^j \,,
\end{split} 
\label{Lgb_split}
\ee
where  $\nabla_\pm x^i \equiv D_\pm x^i + A_a{}^i D_\pm y^a$.
We can rewrite \eqref{Lgb_split} in terms of $\Odd$ symmetric variables using the doubled formalism of Hull \cite{Hull:2004in, Hull:2006va}. 
Firstly, we introduce dual coordinates $\tilde x$ and define
\be
\XX^M = \begin{pmatrix} x^i \\ \tilde x_i \end{pmatrix} \,.
\ee
The geometric data contained in $g_{ij}$ and $b_{ij}$ can be encoded instead in an $\Odd$ generalised metric\footnote{For want of symbols we reuse $\eta, {\cal H}, {\cal P}$ as doubled quantities but now for $\Odd$ rather than $\ODD$; hopefully context serves to disambiguate. }
\be
\mathcal{H}_{MN} = \begin{pmatrix}
g_{ij} - b_{ik} g^{kl} b_{lj} & - b_{ik} g^{kj}
\\
g^{ik} b_{kj} & g^{ij} 
\end{pmatrix} \,,
\label{cosetmetric}
\ee
which obeys $\eta^{-1} \mathcal{H} \eta^{-1} \mathcal{H} = 1$ with $\eta$ the $\Odd$ invariant.
We further introduce $\mathbb{A}_a{}^M = ( A_a{}^i \, A_{ai})$
which serves as a connection for the doubled space with coordinates $\XX^M$, if we view it as a fibration over a base with coordinates $y^a$.
Defining $\nabla_\pm \XX^M \equiv D_\pm \XX^M + \mathbb{A}_a{}^M D_\pm y^a$, the doubled worldsheet Lagrangian is:
\begin{gather}
\begin{split} 
\mathcal{L}_{\text{Hull}}
& =
(g_{ab} +b_{ab} ) D_+ y^a D_-y^b + \tfrac{1}{2} \mathcal{H}_{MN} \nabla_+ \XX^M \nabla_- \XX^N 
\\ & \qquad + \tfrac{1}{2} \mathbb{A}_{a}{}^N \eta_{NM} ( \nabla_+ \XX^M D_- y^a - D_+ y^a \nabla_-\XX^M )
- \tfrac{1}{2} \Omega_{MN} D_+ \XX^M D_- \XX^N \,.
\end{split} 
\label{eq:Ldoubled}
\end{gather}
The very final term is a total derivative, with
\begin{equation}
\Omega = 
\begin{pmatrix}
0&\I1_{d}\\
-\I1_{d}&0
\end{pmatrix} 
\,.
\end{equation}
This can be interpreted as a symplectic form in the doubled space.

To avoid introducing new degrees of freedom, the Lagrangian eq.~(\ref{eq:Ldoubled}) is supplemented by the constraints
\be
{\cal P}_-\nabla_+\,\IX={\cal P}_+\nabla_-\,\IX=0\,,\label{conHull}
\ee
where we recall the projectors 
\begin{equation}
{\cal P}_\pm = \tfrac{1}{2} \left( \I1 \pm  \eta^{-1} \mathcal{H} \right) \label{props} \,. 
\end{equation}
These constraints can be viewed as a (twisted) self-duality condition obeyed by the scalars $\XX^M$ on the worldsheet whose solution determines half of the coordinates $\{x,\tilde x\}$ in terms of the other half. 

This elimination can be done explicitly at the level of the action by gauging the shift symmetry in the dual coordinates. 
This involves introducing an auxiliary worldvolume gauge field which we write as $C_\pm^M = ( 0 , \tilde C_{\pm i})$.
The gauged action has the following formally $O(d,d;\IR)$ covariant form:
\begin{gather}
\begin{split} 
\mathcal{L}_{\text{Hull gauged}}
& =
E_{ab} D_+ y^a D_-y^b + \tfrac{1}{2} \mathcal{H}_{MN} (\nabla_+ \XX^M + C_+^M) (\nabla_- \XX^N + C_-^N)
\\ & \qquad + \tfrac{1}{2} \mathbb{A}_{a}{}^N \eta_{NM} (( \nabla_+ \XX^M + C_+^M) D_- y^a - D_+ y^a (\nabla_-\XX^M + C_-^M) )
\\& \qquad + \tfrac{1}{2} \eta_{MN}(C_+^M D_- \XX^N - D_+ \XX^M C_-^N )
- \tfrac{1}{2} \Omega_{MN} D_+ \XX^M D_- \XX^N \,.
\end{split} 
\label{eq:Ldoubledgauged}
\end{gather}
The equations of motion of $\tilde C_{\pm i}$ are algebraic and equivalent to the constraints.
They allow the $\tilde x$ coordinates to be completely eliminated from \eqref{eq:Ldoubledgauged}, giving the original action \eqref{Lgb_split}.

\subsection{The doubled $\mathcal{N}=(2,2)$ model} 

In this section we demonstrate how to formulate the doubled sigma model directly in $\mathcal{N} = (2,2)$ superspace.

We focus first on the simpler case of BiLP geometries, i.e. generalized K\"ahler manifolds with commuting complex structures. 
This allows for a description solely in terms of chiral and twisted chiral fields.
We summarise our $\mathcal{N}=(2,2)$ superspace conventions in appendix \ref{convs}.
We denote chiral fields $z^\alpha$, $\alpha=1,\dots , n_c$ and twisted chiral fields by $w^\mu$, $\mu=1,\dots n_t$. We let $d=n_c+n_t$.
These fields are complex: we introduce their real and imaginary parts by writing
\begin{equation}
z^\alpha =\frac 1 2 \big( y^\alpha +i\, x^\alpha \big)\,,\qquad w^\mu =\frac 1 2 \big( y^\mu +i\, x^\mu \big)\,.
\end{equation}
The $N=(2,2)$ Lagrangian describing a BiLP geometry is encoded in a single function, the generalized K\"ahler potential, $V(z,w,\bar z, \bar w)$, with the action 
\begin{equation}
S = \int {\rm d}^2\sigma {\rm d}^2\theta {\rm d}^2 \hat \theta \, V(z,w,\bar z, \bar w) \,.
\end{equation}
We now restrict our attention to geometries admitting isometries, with the assumption that the imaginary parts of the chiral and twisted chiral superfields correspond to adapted coordinates for these geometries, i.e. the geometry is independent of the coordinates $(x^\mu,x^\alpha)$.
We can accordingly take the generalized K\"ahler potential to be independent of the isometry directions, thus $V=V(z+ \bar z, w+\bar w)$.

Given a generalized K\"ahler potential of this form, we can define T-dual potentials by Legendre transforming with respect to any of the combinations $z+\bar z$ or $w+\bar w$.
Our construction requires the partial T-dual potentials, $V^{(c)}(z+\bar z, \tilde z+\bar{\tilde{ z}})$ and 
$V^{(t)}(\tilde w+\bar{\tilde{ w}},w+\bar w)$, obtained as Legendre transformations of $V$ with respect to all the twisted chiral and chiral fields respectively
\be
\begin{split} 
V^{(c)}(z+\bar z, \tilde z+\bar{\tilde{ z}}) &= V(z+\bar z, p)+p^ \mu( \tilde z{}_ \mu +\bar{\tilde{ z}}_{ \bar \mu })\,, 
\\
V^{(t)}(\tilde w+\bar{\tilde{ w}},w+\bar w) &= V(q, w+\bar w)-q^ \alpha ( \tilde w{}_ \alpha +\bar{\tilde{ w}}_{ \bar \alpha })\,.
\end{split} 
\label{V12}
\ee
This Legendre transformation introduces $n_t$ (dual) chiral fields $\tilde z_\mu$ and $n_c$ (dual) twisted chiral fields $\tilde w_\alpha$.
We again write the real and imaginary parts as
\be
\tilde w_\alpha = \tfrac12 ( \tilde y_\alpha + i \tilde x_\alpha) \,,\quad
\tilde z_\mu = \tfrac12 (\tilde y_\mu + i \tilde x_\mu ) \,.
\ee
We then define the \emph{doubled-generalised K\"ahler potential} $\IV$ as the sum of the two partial duals:
\begin{eqnarray}
\IV(z,w,\tilde w, \tilde z)=\tfrac 1 2 \big( V^{(c)}(z+\bar z, \tilde z+\bar{\tilde{ z}})+ V^{(t)}(\tilde w+\bar{\tilde{ w}},w+\bar w)
\big)\,,\label{doubledpot}
\end{eqnarray}
We can then define the action for our doubled $\mathcal{N}=(2,2)$ sigma model:
\begin{equation}
\int  {\rm d}^2\sigma {\rm d}^2\theta {\rm d}^2 \hat \theta \, \mathbb{V}(z,w,\tilde w, \tilde z)\,.
\end{equation}
This action is supplemented with the following constraints, which result from the Legendre transformations:
\begin{eqnarray}
&&V_\alpha (z+\bar z, w+\bar w)= (\tilde w +\bar{\tilde w})_ \alpha\,, \nonumber\\
&& V_ \mu (z+\bar z, w+\bar w)= -(\tilde z +\bar{\tilde z})_ \mu \,.\label{cvcon1}
\end{eqnarray}
or concisely in terms of the real parts of superfields as\footnote{Henceforth, we use the notation $V_\mu \equiv \partial_\mu V$, $V_{\mu\alpha} \equiv \partial_\mu \partial_\alpha V$, etc., where now we consider the derivatives with respect to $y_\alpha$ and $y_\mu$.}
\be
V_\alpha (y)  = {\tilde y}_ \alpha\,, 
\quad
V_ \mu (y) = -{\tilde y}_ \mu \,.
\label{cony}
\ee
\subsection{Reduction from $\mathcal{N}=(2,2)$ doubled model to $\mathcal{N}=(1,1)$ doubled model}

Here we verify that the doubled $\mathcal{N}=(2,2)$ model is sensible in that when it is reduced to $\mathcal{N}=(1,1)$ language by performing the integration over the second additional superspace coordinate, $\int d^2 \hat \theta \equiv \hat D_+ \hat D_-$,    we recover the   $\mathcal{N}=(1,1)$ doubled sigma model introduced above.

Let us first consider the usual $\mathcal{N}=(1,1)$ description of the BiLP geometry introduced in \eqref{SVorig} but now imposing isometries in the $x^\alpha, x^\mu$ directions (the imaginary parts of chiral and twisted chiral multiplets respectively).   The   $\mathcal{N}=(1,1)$ Lagrangian that one finds can be written as $\mathcal{L} = \mathcal{L}_x + \mathcal{L}_y$ with
\be
\begin{split} 
\mathcal{L}_x &=g_{ \alpha \beta }D_+x^ \alpha D_- x^ \beta +g_{ \mu \nu }D_+x^\mu D_-x^\nu +b_{ \alpha \nu }D_+ x^ \alpha D_- x^ \nu +b_{ \mu \beta }D_+x^\mu D_-x^ \beta \,,\\
\mathcal{L}_y & = g_{ \alpha \beta }D_+y^ \alpha D_- y^ \beta +g_{ \mu \nu }D_+y^\mu D_-y^\nu +b_{ \alpha \nu }D_+ y^ \alpha D_- y^ \nu +b_{ \mu \beta }D_+y^\mu D_-y^ \beta\,, 
\end{split}
\label{origL}
\ee
where the metric and $B$-field are derived as second derivatives (implicitly understood to be with respect to the coordinates $y$) of the potential:
\be
g_{ \alpha \beta }  =+V_{ \alpha \beta }\,,\quad g_{ \mu \nu   }=-V_{ \mu  \nu  }\,, \quad
b_{ \alpha \nu  }  =-V_{ \alpha \nu  }\,,\quad b_{ \mu \beta    }=+V_{ \mu  \beta   }\,.
\ee
Observe that the fibre and spectator terms have identical metric and $B$-field components.
The $B$-field terms in $\mathcal{L}_y$ are in fact a total derivative. Discarding these terms, the doubled $\mathcal{N}=(1,1)$ sigma model \eqref{eq:Ldoubled} derived from \eqref{origL} has the Lagrangian
\be
\mathcal{L} 
= \tfrac{1}{2} \mathcal{H}_{MN} D_+ \XX^M D_- \XX^N 
+ g_{ \alpha \beta }D_+y^ \alpha D_- y^ \beta +g_{ \mu \nu }D_+y^\mu D_-y^\nu \,.
\label{targetdoubled}
\ee
The $O(d,d)$ generalised metric appearing here has the functional form of \eqref{cosetmetric} with the specific $d$-dimensional metric and B-field:
\be
g_{ij} \equiv \begin{pmatrix} g_{\alpha \beta} & 0 \\ 0 & g_{\mu\nu} \end{pmatrix} \,,\quad
b_{ij} \equiv \begin{pmatrix} 0 & b_{\alpha \nu} \\ b_{\mu \beta} & 0 \end{pmatrix} \,.
\label{GB}
\ee
The doubled coordinates are $\XX^M = ( x^i,\tilde x_i)$ with $x^i=(x^\alpha,x^\mu)$, $\tilde x_i=(\tilde x_\alpha,\tilde x_\mu)$, obeying the constraint \eqref{conHull} which is equivalent to 
\be
D_\pm \tilde x_i = \pm (g\mp b)_{ij} D_\pm x^j \,.
\label{xtildexDUAL}
\ee

Now we wish to show how the $\mathcal{N}=(2,2)$ doubled model arising from the doubled generalised K\"ahler potential eq.~\eqref{doubledpot}  reproduces the doubled model in $\mathcal{N}=(1,1)$ superspace.
We again have to perform the integration $\int d^2 \hat \theta \equiv \hat D_+ \hat D_-$, and therefore we need the double derivatives of the doubled potential \eqref{doubledpot}.
We can easily obtain these derivatives, and relate them to derivatives of the original potential $V$, using properties of the Legendre transformation defining $V^{(c)}$ and $V^{(t)}$ in \eqref{V12}.  For instance, by taking a total derivative of the defining relation
  \be
V^{(c)}(y^\alpha, \tilde y_\mu) = V(y^\alpha,y^\mu)+y^\mu \tilde y_\mu  
\,,  
\ee 
we have the conditions 
\be  
    V_\mu  = - \tilde{y}_\mu \, , \quad
    V_\alpha  = V^{(c)}_\alpha \, ,  \quad   V^{(c)}{}^\mu  = y^\mu \, . 
\ee
A subsequent total derivative of each these conditions produces
\be 
\begin{aligned} 
    V_{\mu\nu} dy^\nu +   V_{\mu\alpha } dy^\alpha     &= - d\tilde{y}_\mu \, , \\  
    V_{\alpha \nu } dy^\nu + V_{\alpha \beta} dy^\beta    &= V^{(c)}_\alpha{}^\nu d\tilde{y}_\nu +V^{(c)}_{\alpha \beta} dy^\beta   \, , \\  V^{(c)}{}^{\mu \nu} d\tilde{y}_\nu +  V^{(c)}{}^{\mu }{}_\alpha  dy^\alpha    &= dy^\mu \, . 
    \end{aligned} 
\ee
Solving simultaneously returns the result that 
\be
\begin{split} 
V^{(c)}_{ \alpha \beta } &=V_{ \alpha \beta }-V_{ \alpha \mu } \big( V^{-1}\big)^{ \mu \nu }V_{\nu \beta }\,,\quad
V^{(c)}_ \alpha {}^\nu =V^{(c)}{}^ \nu {}_ \alpha =- V_{ \alpha \mu } \left( V^{-1}\right)^{ \mu \nu }\,, 
\\
V^{(c)}{}^{ \mu \nu } & = -\big( V^{-1}\big)^{ \mu \nu }\, , 
\label{ders1}
\end{split}
\ee
in which $(V^{-1})^{\mu\nu}$ means the matrix inverse of $V_{\mu\nu}$.  The same manipulations for the dualisation to twisted chirals gives 
\be
\begin{split} 
V^{(t)}_{ \mu \nu }&=V_{ \mu \nu }-V_{ \mu \alpha } \big( V^{-1}\big)^{ \alpha \beta }V_{ \beta \nu }\,,\quad
V^{(t)}_ \mu {}^ \beta =V^{(t)}{}^ \beta {}_ \mu = V_{ \mu \alpha } \big( V^{-1}\big)^{ \alpha \beta }\,, 
\\
V^{(t)}{}^{ \alpha \beta } & = -\big( V^{-1}\big)^{ \alpha \beta }\,.\label{ders2}
\end{split}
\ee
Invoking eqs.  \eqref{ders1} and  \eqref{ders2} we immediately observe that these derivatives correspond to the components of the generalised metric:
\be
{\mathcal{H}}_{MN} = \begin{pmatrix}
V^{(c)}_{\alpha \beta} & 0 &0 & V^{(c)}_{\alpha}{}^\mu \\ 
0 & - V^{(t)}_{\mu \nu} & - V^{(t)}_{\mu}{}^\alpha & 0 \\
0 & - V^{(t)\, \alpha}{}_\nu & - V^{(t) \, \alpha \beta} & 0 \\
V^{(c)\,\mu}{}_\alpha & 0 &0 & V^{(c) \, \mu \nu }
 \end{pmatrix}\, . 
\ee
Notice that this is not simply the Hessian of the doubled potential $\mathbb{V}(\mathbb{Y})$ with respect to the doubled coordinates $\mathbb{Y}^M=(y^ \alpha\ y^ \mu \ \tilde y_ \alpha\ \tilde y_\mu )$ but instead contains some additional signs. These extra signs arise  arise,  when passing from ${\cal N}=(2,2)$ to $\mathcal{N}=(1,1)$,  from the complex structures to which the chiral and twisted chiral fields are adapted.

Consequently,  the doubled model in $\mathcal{N}=(1,1)$ superspace is given by  
\begin{eqnarray}
{\cal L}=\frac 1 2\, D_+\IX^M {\cal H}_{MN} D_-\IX^N+ \frac 1 2\, D_+\IY^M {\cal H}_{MN} D_-\IY^N\,,\label{duallag1}
\end{eqnarray}
where $\mathbb{X}^M=(x^ \alpha\ x^ \mu \ \tilde x_ \alpha\ \tilde x_\mu )$ and $\mathbb{Y}^M=(y^ \alpha\ y^ \mu \ \tilde y_ \alpha\ \tilde y_\mu )$ and $ {\cal H}_{MN}$ as given in equation \eqref{cosetmetric} in terms of the metric and $B$-field \eqref{GB}.
A priori the ``spectator'' part of the Lagrangian looks quite different from \eqref{eq:Ldoubled} as it is also, formally, doubled.    However, we need to also consider the constraints \eqref{cvcon1} or \eqref{cony} which in   $\mathcal{N}=(1,1)$ superspace derivatives of \eqref{cony} imply
\begin{gather}
\begin{split}
D_\pm \tilde y_\alpha & = V_{\alpha \beta} D_\pm y^\beta + V_{\alpha \mu} D_\pm y^\mu \,,\\
D_\pm \tilde y_\mu & = - V_{\mu\nu} D_\pm y^\nu - V_{\mu \alpha} D_\pm y^\alpha \,,
\end{split} 
\label{Dcony}
\end{gather}
which can be rewritten using $y^i \equiv (y^\alpha \, y^\mu)$ and $\tilde y_i \equiv (\tilde y_\alpha\,\tilde y_\mu)$ as 
\be
D_\pm \tilde y_i = (g-b)_{ij} D_\pm y^j \,.
\label{ytildeyNOTdual}
\ee
Note this is {\em not} the standard T-duality constraint; left and right movers (i.e. plus and minus derivatives in \eqref{ytildeyNOTdual}) are related in the same fashion whereas in T-duality the transformation of left and right movers differ by a target space local Lorentz rotation (as per discussion in section \ref{transJT}).

Next for the isometry directions we act on \eqref{cony} using $\hat{D}_{\pm}$.
Using the $\mathcal{N}=(2,2)$ superspace constraints for chiral and twisted chiral multiplets, see eq.~\eqref{actuallyuseful}, this gives 
\begin{gather}
\begin{split}
D_\pm\tilde x_ \alpha &=\pm V_{\alpha \beta } (y)D_\pm x^ \beta + V_{\alpha \nu } (y)D_\pm x^\nu\,, \\
D_\pm \tilde x_ \mu &= \mp  V_{\mu \nu } (y)D_\pm x^\nu  -V_{\mu \beta } (y)D_\pm x^ \beta \, , 
\end{split}
\label{conal2}
\end{gather}
which can be rewritten using $x^i \equiv (x^\alpha \, x^\mu)$ and $\tilde x_i \equiv (\tilde x_\alpha\,\tilde x_\mu)$ as
\be 
D_\pm \tilde{x}_i = \pm (g_{ij} \mp b_{ij}) D_\pm x^j \, ,
\ee
i.e. exactly recovering the anticipated  chirality constraints \eqref{conHull} of the doubled $\mathcal{N}=(1,1)$ string in the form \eqref{xtildexDUAL}.


We now return to the Lagrangian \eqref{duallag1}. 
Although it would be inconsistent to use the constraints \eqref{conal2} in the action, we are allowed insert \eqref{Dcony} or \eqref{ytildeyNOTdual} in order to eliminate the $\tilde y$ coordinates in terms of the $y$ coordinates. 
This is because these are related by the (local) field redefinition \eqref{cony} whereas the relationship between $x$ and $\tilde x$ has the interpretation of a chirality constraint.
Once we eliminate the coordinates $\tilde y$ from \eqref{duallag1} we obtain exactly the $\mathcal{N}=(1,1)$ doubled Lagrangian \eqref{eq:Ldoubled} .

Furthermore, we can also directly obtain both the $x$ and $y$ equations of motion by taking further derivatives of the constraint in the form \eqref{conal2}.
For instance, the equation of motion of $x^\alpha$ follows from 
\begin{gather}
\begin{split} 
0 & = D_- D_+\tilde x_\alpha + D_+ D_- \tilde x_\alpha 
\\ & = D_- ( V_{\alpha \beta } (y)D_+x^ \beta + V_{\alpha \nu } (y)D_+ x^\nu ) 
+ D_+(-V_{\alpha \beta } (y)D_-x^ \beta + V_{\alpha \nu } (y)D_- x^\nu) \,,
\end{split}
\end{gather}
while acting again with $\hat D_\pm$ we have for example
\begin{gather}
\begin{split} 
\hat D_- D_+ \tilde x_\alpha = D_+ D_- \tilde y_\alpha 
& = \hat D_- ( V_{\alpha \beta } (y)D_+x^ \beta + V_{\alpha \nu } (y)D_+ x^\nu ) \\
& = - V_{\alpha \beta \gamma} D_- x^\gamma D_+ x^\beta 
+ V_{\alpha \beta \mu} D_- x^\mu D_+ x^\beta 
- V_{\alpha \beta} D_+D_- y^\beta
\\ & \quad
- V_{\alpha \nu \gamma} D_- x^\gamma D_+ x^\nu 
+ V_{\alpha \mu \nu } D_- x^\mu D_+ x^\nu
+ V_{\alpha \mu} D_+ D_- y^\mu \,,
\end{split}
\end{gather}
and after substituting in for $\tilde y_\alpha$ using \eqref{cony} this can be checked to reproduce the $y^\alpha$ equation of motion.


A final comment concerns the topological term of the $\mathcal{N}=(1,1)$ doubled string \cite{Hull:2004in, Hull:2006va}.
Whilst this term breaks $\Odd$ symmetry, its inclusion is important in gauging and to obtain correct answers at the quantum level.
Here we point out that such a term arises in $\mathcal{N}=(2,2)$ double model from  the following quantity: 
\be
\varpi (y,\tilde y)
= \tfrac12 (  y^\alpha \tilde y_\alpha  -   y^\mu \tilde y_\mu) \,.
\label{topterm}
\ee
One has that $\hat D_+ \hat D_- \varpi$ gives (without dropping total derivatives) 
\be
\mathcal{L}_\varpi = \tfrac12 \Omega_{MN} D_+ \XX^M D_- \XX^N + \tfrac12 \Omega_{MN} \YY^M D_-D_+  \YY^N\,,
\label{Lvarpi}
\ee
where the constant antisymmetric matrix $\Omega_{MN}$ has non-zero components $\Omega^\alpha{}_\beta  = + \delta^\alpha_\beta = - \Omega_\beta{}^\alpha$, $\Omega^\mu{}_\nu = + \delta^\mu{}_\nu=-\Omega_\nu{}^\mu$.
Each term in \eqref{Lvarpi} is a total derivative.
The doubled potential $\mathbb{V}$ is such that $V = \mathbb{V} + \varpi$. Hence adding or subtracting the total derivative $\varpi$ in $\mathcal{N}=(2,2)$ superspace, and using the constraint, breaks the duality symmetry.

In the $\mathcal{N}=(1,1)$ doubled approach, making different choices of $\Omega_{MN}$ corresponds to picking different choices of duality frame.
Here we can implement this by making different choices of $\varpi$.
For instance, flipping the sign allows us to pick out the totally T-dual potential 
\be
\widetilde V \equiv \mathbb{V} - \varpi 
= V - y^\alpha \tilde y_\alpha + y^\mu \tilde y_\mu \,.
\label{tildeV}
\ee
An alternative choice such as $\varpi^{(c)}
= \tfrac12 (y^\alpha \tilde y_\alpha + y^\mu \tilde y_\mu)$ gives back $V^{(c)} = \mathbb{V} + \varpi^{(c)}$. These different choices of $\varpi$ (with different choices of sign) lead to different $\Omega_{MN}$ related by Buscher T-duality.

\subsection{$\Odd$ and the ${\mathcal N}=(2,2)$ doubled model} 
\label{oddtwotwo}

\subsubsection{The constraint as a Lagrangian}
Critical to the $\mathcal{N}=(2,2)$ doubled model was the constraint of eq.~\eqref{cony} that encodes not only the duality relations between coordinates $x$ and $\tilde{x}$ interchanged by T-duality on the fibre but also an un-doubling of the formally doubled base coordinates $y$ and $\tilde{y}$.   To get a better understanding of this, and how T-duality acts on the doubled model,  we  first formulate the constraint \eqref{cony} in a more democratic form. 
From the definitions of $V^{(c)}$ and $V^{(t)}$ in \eqref{V12}, it follows that 
\be
d \mathbb{V}
= \tfrac12 ( \tilde y_\alpha dy^\alpha - y^\alpha d \tilde y_\alpha
- \tilde y_\mu d y^\mu + y^\mu d \tilde y_\mu ) \,.
\label{dVV}
\ee
In terms of $\mathbb{Y}^M=(y^ \alpha\ y^ \mu \ \tilde y_ \alpha\ \tilde y_\mu )$ we can recast this expression as  
\be
\mathbb{V}_M \equiv \frac{\partial \mathbb{V} }{\partial \mathbb{Y}^M}  = \Theta_{M N} \YY^N \,,
\label{VTheta} 
\ee
with
\be
\Theta_{MN} = \tfrac{1}{2} \begin{pmatrix} 0 & \sigma \\ - \sigma & 0 \end{pmatrix} 
\,,\quad
\sigma \equiv \begin{pmatrix} \I1_{n_c} & 0 \\ 0 & - \I1_{n_t} \end{pmatrix} \,.
\label{defThetasigma}
\ee
The object $\Theta_{MN}$ is suggestive of a symplectic structure. 
Indeed, we could regard \eqref{dVV} as being the statement that $d \mathbb{V} = \vartheta$ with 
$ 
\vartheta \equiv \Theta_{MN} \YY^N d \YY^{M}  
$
a potential for a symplectic form $\omega = - d \vartheta$   given by
\be
\omega = \Theta_{MN} d \YY^M \wedge d \YY^N 
=  d \tilde y_\alpha \wedge d y^\alpha - d \tilde y_\mu \wedge d y^\mu \,.
\label{conomega}
\ee
The constraint \eqref{dVV} implies that we have $\omega=0$ on the constraint surface.
Hence, the doubled model is defined on a Lagrangian submanifold of the `doubled' base space with coordinates $\YY^M$ and symplectic form $\omega$.

\subsubsection{Doubled and over-doubled complex structures}

Although we have been working in  coordinates manifestly adapted to both isometries and complex structures, it is helpful to note that the over-doubled space itself inherits complex structures:   
\be
\hat D_\pm \begin{pmatrix} \XX \\ \YY \end{pmatrix} 
= \mathbb{J}_\pm D_\pm \begin{pmatrix} \XX \\ \YY \end{pmatrix} \,,
\label{overdoubledhatD}
\ee
with   $\mathbb{J}_\pm$ such that 
\be
\mathbb{J}_+ = \begin{pmatrix} 0 & j_+ \\ 
- j_+ & 0 
\end{pmatrix} 
\,,\quad
\mathbb{J}_- = \begin{pmatrix} 0 & j_- \\ 
- j_- & 0 
\end{pmatrix} 
\label{overdoubledJ}
\ee
where for $n_c$ chiral, $n_t$ twisted chiral, $n_c$ dual twisted chiral and $n_t$ dual chiral fields, with $n_c+n_t=d$, we have that $j_+$ and $j_-$ are $2d $ matrices given by:
\be
j_+ = \I1_{2d} \,,\quad
j_- = \begin{pmatrix}
\sigma & 0 \\  0 & -\sigma
\end{pmatrix} \,,
\ee
with $\sigma$ as in \eqref{defThetasigma}.
A simple calculation then shows that the $\mathcal{N}=(1,1)$ Lagrangian that follows on carrying out the integration over the extra $\mathcal{N}=(2,2)$ fermionic coordinates has the form
\be
\mathcal{L}
= \mathbb{V}_{KL} j_+^K{}_M j_-^K{}_N D_+ \XX^M D_- \XX^N 
+ \mathbb{V}_{MK} j_-^K{}_L j_+^L{}_N D_+ \YY^M D_- \YY^N \,,
\ee
where the derivatives of $\mathbb{V}(\YY)$ are always with respect to $\YY^M$.
Comparing with \eqref{duallag1} this implies that the $\Odd$ generalised metric which appears in both terms is given by 
\be
\mathcal{H}_{MN} = 2 \mathbb{V}_{KL} j_+^K{}_M j_-^L{}_N =  2 \mathbb{V}_{MK} j_-^K{}_L j_+^L{}_N \,.
\ee
Given that $j_+^M{}_N = \delta^M_N$, these two expressions are indeed equal.

We can further make an identification 
\be
\Theta_{MN} =  \frac{1}{2} j_-^K{}_M j_+^L{}_K \eta_{L N} \,,
\ee
using $\eta$ the $\Odd$ structure. 
We then note that taking an exterior derivative of \eqref{VTheta} implies 
\be
\mathbb{V}_{MN} d \YY^N = \Theta_{MN} d \YY^N 
\Rightarrow d \YY^M = \eta^{MN} \mathcal{H}_{NK} d \YY^K
\ee
which reproduces the constraint \eqref{ytildeyNOTdual} (and to reiterate for emphasis, the absence of a Hodge star in this relation is in contradistinction to that of the T-duality relation obeyed by the $\mathbb{X}$).

\subsubsection{Transformation rules} 
\label{OddRules} 
Let us now look at the T-duality transformation properties of our model.
From the $\mathcal{N}=(1,1)$ case, we know that ${\cal O} \in \Odd$   acts on $\IX$ as,
\be
\IX \rightarrow \IX'= {\cal O}^{-1}\,\IX\,.\label{altX}
\ee
We require that these transformations respect the $\mathcal{N}=(2,2)$ supersymmetry transformations, which implies we should also have an action of ${\cal O}$ on $\IY$,
\be
\IY \rightarrow \IY'= {\cal O}^{-1}\,\IY\,. \label{altY}
\ee
This is accompanied by a transformation of the complex structures $\mathbb{J}_\pm$ of \eqref{overdoubledJ}:
\be
\mathbb{J}_\pm \rightarrow \mathbb{J}_\pm' = \begin{pmatrix} \mathcal{O}^{-1} & 0 \\ 0 & \mathcal{O}^{-1}  \end{pmatrix} 
\mathbb{J}_\pm \begin{pmatrix} \mathcal{O} & 0 \\ 0 & \mathcal{O}  \end{pmatrix} \,.
\label{JJprime}
\ee
In order to reproduce the behavior of the doubled action in $\mathcal{N}=(1,1)$ superspace under $\Odd$ transformation we then have to require that the doubled potential transforms as a scalar,
\be
\IV(\IY) \rightarrow \IV'(\IY')=\IV(\IY)\,.\label{altV}
\ee
We must also require that the constraint transformations covariantly, in particular we need
\be
\Theta'={\cal O}^T\Theta {\cal O}\,. 
\ee
The T-duality invariance of the doubled model means that although we initially defined the doubled potential starting from a BiLP geometry, it can describe more complicated cases. 
For example, as we saw at the end of section \ref{transJT}, T-duality transformations can take us from a BiLP geometry with commuting complex structures to a geometry with non-commuting complex structures. The latter would normally have to be described with semi-chiral superfields.
However, we can alternatively describe the geometry using appropriate combinations of our original doubled chiral and twisted chiral superfields, and thereby avoid introducing semi-chirals.
We will demonstrate this explicitly in an example in section \ref{semi}.

\subsubsection{Transformation of complex structures in detail} 
\label{JJworks}

In order to verify that the above transformation rules are sensible, we should consider more carefully the implications of the transformation \eqref{JJprime} of the complex structures of  \eqref{overdoubledJ}.
These are initially given in adapted coordinates, and their geometric nature is unclear. 
Despite this, we can show that their transformation \eqref{JJprime} under $\Odd$, together with the transformation of the constraint, allow us to recover the correct transformation of the usual complex structures $J_\pm$ as derived in section \ref{transJT}.

Firstly, we note that as we start with $\mathbb{J}_+$ of the form \eqref{overdoubledJ} with $j_+ = \I1_{2d}$, the $\Odd$ transformation \eqref{JJprime} always preserves $\mathbb{J}_+' = \mathbb{J}_+$. 
However, generically $\mathbb{J}_-$ will transform non-trivially and lead to a very non-standard complex structure which in particular mixes physical and dual superfields. 
The correct prescription to recover sensible complex structures acting solely on the physical superfields is to apply the constraint to eliminate the dual ones. Then, after applying $\Odd$ transformation and the constraint, one obtains new complex structures $J_\pm'$, with the special feature that $J_+'$ is always in `diagonal' (w.r.t. a complex basis) form. 
This corresponds to doing a standard $\Odd$ transformation as in section \ref{transJT} followed by a coordinate transformation of the non-isometric directions in order to bring $J_+'$ into the desired form.

We now prove this. 
Let's start with the complex structures corresponding to our $2d$-dimensional BiLP geometry with $n_c$ chiral superfields and $n_t$ twisted chiral superfields. 
In adapted coordinates for the $d$ isometry directions we can write (as in appendix \ref{Jadapt})
\be
J_+ = \begin{pmatrix} 0 & \I1_d \\ - \I1_d & 0 \end{pmatrix} \,,\quad
J_- = \begin{pmatrix} 0 & \sigma \\ - \sigma & 0 \end{pmatrix} \,.
\ee
The metric and $B$-field are meanwhile given by
\be
G+B=\begin{pmatrix} g+b & 0 \\ 0 & g+b \end{pmatrix} \,,\quad
g = \begin{pmatrix} g_{cc} & 0 \\ 0 & g_{tt} \end{pmatrix} \,,\quad
b = \begin{pmatrix} 0 & b_{ct} \\ - b_{ct} & 0 \end{pmatrix} \,,
\ee
where we schematically exhibit the non-zero metric and $B$-field components in chiral and twisted chiral directions.

The T-duality transformation rule \eqref{jstrsf} implies that under an $\mathcal{O} \in \Odd \subset \ODD$ transformation with
\be
\mathcal{O}
=\begin{pmatrix} \MA & \MC \\ \MB & \MD \end{pmatrix}\,,
\ee
we obtain new complex structures
\be
J_+' = \begin{pmatrix} 0 & O_+ \\ - O_+^{-1} & 0 \end{pmatrix} 
\,,\quad
J_-' = \begin{pmatrix} 0 & O_- \sigma \\ - \sigma O_-^{-1} & 0 \end{pmatrix} \,,
\ee
where
\be
O_+ = \MC^t (g-b) + \MD^t \,,\quad
O_- = - \MC^t (g+b) + \MD^t \,.
\ee
We `diagonalise' $J_+'$ by using a coordinate transformation $y \rightarrow y'(y)$ such that $\partial y'/\partial y = O_+$. 
This leads to:
\be
J_+' = \begin{pmatrix} 0 & \I1_d \\ - \I1_d & 0 \end{pmatrix} 
\,,\quad
J_-' = \begin{pmatrix} 0 & O_-\sigma O_+^{-1}  \\ - O_+ \sigma O_-^{-1} & 0 \end{pmatrix} \,.
\ee
We then have
\be
\begin{split} 
\hat D_- x' & = O_- \sigma O_+^{-1} D_- y'
= ( - \MC^t (g+b) + \MD^t ) \sigma (\MC^t (g-b) + \MD^t)^{-1} D_- y' \,,\\
\hat D_- y' & = - O_+ \sigma O_-^{-1} D_- x'
= -  (\MC^t (g-b) + \MD^t) \sigma ( - \MC^t (g+b) + \MD^t )^{-1} D_- x' \,.
\end{split}
\label{targetxy}
\ee
We wish to reproduce these relationships from our $\mathcal{N}=(2,2)$ doubled model. 
According to \eqref{JJprime}, we obtain new complex structures $\JJ'_\pm$ with $\JJ'_+=\JJ_+$ and
\be
\JJ'_- = \begin{pmatrix} 
0 & j'_- \\
- j'_- & 0 
\end{pmatrix} 
\,,\quad
j'_- =
\begin{pmatrix}
\MD^t \sigma \MA - \MC^t \sigma \MB & 
\MD^t \sigma \MC - \MC^t \sigma \MD \\
\MB^t \sigma \MA - \MA^t \sigma \MB & 
\MB^t \sigma \MC - \MA^t \sigma  \MD 
\end{pmatrix} \,.
\ee
Hence we have
\be
\begin{split} 
\hat D_- x' & = (\MD^t \sigma \MA - \MC^t \sigma \MB ) D_- y' 
+ (\MD^t \sigma \MC - \MC^t \sigma \MD) D_- \tilde y' \\
\hat D_- y' & = -(\MD^t \sigma \MA - \MC^t \sigma \MB ) D_- x' 
- (\MD^t \sigma \MC - \MC^t \sigma \MD) D_- \tilde x'\,,
\end{split} 
\label{naivexprimeyprime}
\ee
We eliminate the dual coordinates from \eqref{naivexprimeyprime} using the transformed constraints:
\be
\begin{split} 
D_- \tilde y'& = (g'-b') D_- y' 
= ( \MA^t (g-b) + \MB^t ) O_+^{-1} D_-y' \,,\\
D_- \tilde x'& = -(g'+b') D_- y' 
= -( \MA^t (g+b) - \MB^t ) O_-^{-1} D_-y' \,,
\end{split} 
\label{constraintsprime}
\ee
using \eqref{bes1} and \eqref{bes2} to express the transformation of the geometry.
Starting with $\hat D_- x'$, we can thus write:
\be
\begin{split}
\hat D_- x' & = \left( (\MD^t \sigma \MA - \MC^t \sigma \MB )
O_+
+ (\MD^t \sigma \MC - \MC^t \sigma \MD)  ( \MA^t (g-b) + \MB^t ) \right) O_+^{-1} D_- y' \\
& = \Big(
\left( \MD^t \sigma ( \MA \MC^t + \MC\MA^t)- \MC^t \sigma (\MB\MC^t +\MD\MA^t) \right) (g-b)
\\ & \qquad\qquad\qquad + \MD^t \sigma ( \MA \MD^t+ \MC\MB^t) - \MC^t \sigma (\MB\MD^t +\MD\MB^t) 
\Big) O_+^{-1} D_- y'
\\ & = 
( - \MC^t \sigma(g-b) \sigma + \MD^t ) \sigma O_+^{-1} D_- y'
\,,
\end{split} 
\ee
where we used the conditions resulting from the fact that the blocks $\MA,\MB,\MC,\MD$ give an $\Odd$ transformation.
Finally, as $\sigma(g-b) \sigma=g+b$ we recover exactly the correct relationship \eqref{targetxy}.
The calculation of $\hat D_- y'$ proceeds similarly.

\section{Examples}
\label{examples}

In this section we illustrate the construction of the doubled potential in a number of examples.
In each case the recipe is as follows:
\begin{itemize}
\item We start with the generalised K\"ahler potential $V(z+\bar z, w+\bar w)$ describing a BiLP geometry with isometries.
\item We construct the totally chiral and twisted chiral potentials, $V^{(c)}(z+\bar z, \tilde z+\bar{\tilde z})$ and $V^{(t)}(\tilde w+\bar{\tilde w}, w+\bar w)$, defined by Legendre transformation in \eqref{V12}.
\item We hence write down the doubled potential $\mathbb{V}$ given by \eqref{doubledpot} and the constraints given by \eqref{cvcon1} or \eqref{cony}.
\end{itemize}

\subsection{Torus with constant B-field}     
\label{torusexample} 

A first simple example is a torus with constant $B$-field.
This can be realised via the following generalised K\"ahler potential:
\begin{equation}
V(z,w) = \tfrac{1}{2} (z+\bar{z})^2 - \tfrac{1}{2} (w+\bar{w})^2 + \lambda (z+\bar{z})(w+\bar{w}) \,,
\end{equation}
Defining $y_1 = z+\bar z$ be the real part of the chiral superfield and $y_2=w+\bar w$ that of the twisted chiral superfield we can rewrite the potential as
\begin{equation}
V(y_1,y_2) = \tfrac12 y_1^2 - \tfrac12 y_2^2 + \lambda y_1 y_2 \,.
\end{equation}
The corresponding geometry is simply:
\be
ds^2 = dx_1^2 + dy_1^2 +  dx_2^2 + dy_2^2 
\,,\quad
B =- \lambda d y_1 \wedge dy_2 - \lambda  d x_1 \wedge dx_2 \,.
\label{torusgeo}
\ee
We now give the doubled $\mathcal{N}=(2,2)$ description.
The dual potentials \eqref{V12} used in the construction of the doubled potential are:
\begin{align}
V^{(c)}(y_1, \ty_2) & = \tfrac12 (1+\lambda^2) y_1^2 + \tfrac12 \ty_2^2 + \lambda y_1 \ty_2 \,,
\\
V^{(t)} (\ty_1, y_2) & = - \tfrac12 \tilde y_1^2 - \tfrac12 (1+\lambda^2) y_2^2 + \lambda \ty_1 y_2 \,,
\end{align}
The doubled potential \eqref{doubledpot} is thus:
\be
2 \VV = \tfrac12 (1+\lambda^2) y_1^2 + \tfrac12 \ty_2^2 - \tfrac12 \tilde y_1^2 - \tfrac12 (1+\lambda^2) y_2^2 + \lambda( y_1 \ty_2+ \ty_1 y_2 )\,.
\ee
and the constraints \eqref{cony} are:
\begin{eqnarray} 
\left( 
\begin{array}{c} 
y_1 \\ \tilde{y}_1 \\ y_2 \\ \tilde{y}_2 
\end{array}\right) 
= 
\begin{pmatrix}
- \tilde{y}_1 + \lambda y_2 \\
(1+\lambda^2) y_1+ \lambda \tilde{y}_2 \\
\tilde{y}_2 + \lambda y_1 \\
-(1+\lambda^2) y_2 + \lambda \tilde{y}_1 
\end{pmatrix}
\,. 
\end{eqnarray} 

\subsection{\texorpdfstring{$\mathrm{SU}(2) \times \mathrm{U}(1)$}{SU(2)xU(1)} } 
\label{SU2U1example}

Another example is provided by the WZW model on $\mathrm{SU}(2)\times \mathrm{U}(1)$, for details of which see \cite{Sevrin:2011mc}. 
In its BiLP description, the generalized K\"ahler potential \cite{Rocek:1991vk} reads, 
\begin{eqnarray} 
V=-\frac 1 2 \, (w +\bar w)^2+ \int^{z+\bar z-w -\bar w}\, dr\,\ln (1+e^r)
\,,\label{KP} 
\end{eqnarray} 
which in our conventions describes the geometry
\be
\begin{split} 
ds^2 & = \frac{1}{1+e^{y_1-y_2}} \left( e^{y_1-y_2} ( dy_1^2 + dx_1^2 ) + dy_2^2 + dx_2^2 \right) \,,\\
B & = \frac{e^{y_1-y_2}}{1+e^{y_1-y_2}} ( dy_1 \wedge dy_2 + dx_1 \wedge dx_2 ) \,.
\end{split} 
\ee
The Legendre dual potentials are:
\begin{eqnarray} 
V^{(c)}(z+\bar z, \tilde z+\bar{\tilde{ z}})&=&\frac 1 2\,(z+\bar z)^2+ 
\int^{z+\bar z -p}dr\,\ln(1+e^{-r})+p( \tilde z +\bar{\tilde{ z}}
-1
(z+\bar z))\nonumber\\ 
&=&\frac 1 2 \, (\tilde z+\bar{\tilde{ z}}
)^2+\int^{\tilde z+\bar{\tilde{ z}}+
z+ \bar z}dr\,\ln(1-e^{-r})\,,\nonumber\\ 
V^{(t)}(\tilde w+\bar{\tilde{ w}},w+\bar w)&=& 
-\frac 1 2 \, (w +\bar w)^2+ \int^{q-w -\bar w}\, dr\,\ln (1+e^r) 
-q ( \tilde w{} +\bar{\tilde{ w}}
) \nonumber\\ 
&=& - \frac 1 2 \, (\tilde w+\bar{\tilde{ w}}+
(w+\bar w))^2-\int^{\tilde w+\bar{\tilde{ w}}
}dr\,\ln(1-e^{-r})\,.\nonumber\\ 
\end{eqnarray} 
Hence we get for the doubled-generalised K\"ahler potential, 
\begin{equation}
\begin{split}
\IV(z+\bar z, w+\bar w, \tilde w+\bar{\tilde{ w}},\tilde z+\bar{\tilde{ z}})& =\frac 1 2 \,\Big( 
\frac 1 2 \, (\tilde z+\bar{\tilde{ z}})^2 - \frac 1 2 \, (\tilde w+\bar{\tilde{ w}}+w+\bar w)^2 
\\ & \quad\qquad+\int^{\tilde z+\bar{\tilde{ z}}-z- \bar z}_{\tilde w+\bar{\tilde{ w}}}dr\,\ln(1-e^{-r}) 
\Big)\,. 
\end{split} 
\end{equation}
The constraints are 
\begin{eqnarray} 
\left( 
\begin{array}{c} 
z +\bar z \\ w+\bar w \\ \tilde w +\bar{\tilde{w}} \\ 
\tilde z +\bar{\tilde{z}} 
\end{array}\right) 
= 
\left( 
\begin{array}{c} 
w+\bar w +\ln(e^{\tilde w +\bar{\tilde{w}}}-1)\\ \ln(e^{\tilde z +\bar{\tilde{z}} }-e^{z+\bar z}) \\ -\ln(1-e^{z+\bar z-\tilde z -\bar{\tilde{z}} }) \\ 
w+\bar w + \tilde w +\bar{\tilde{w}} 
\end{array}\right) 
\,. 
\end{eqnarray}

\subsection{T-fold examples}
\label{tfolds} 
We now consider some more novel examples of $\mathcal{N}=(2,2)$ \emph{non-geometries}. 
The general idea is the following. 
Recall that a generalised K\"ahler potential $V(z,w)$ is defined only up to generalised K\"ahler transformations:
\be
V(z,w) \rightarrow V(z,w) 
+ F(z,w) + \bar F(\bar z, \bar w)
+ G(\bar z,w) + \bar G(z,\bar w) \,,
\ee
for arbitrary functions $F$ and $G$.
The examples we consider below will have this feature when we carry out some transformation of $z$ and $w$ under which the geometry is globally identified. 
We will refer to this as a monodromy of the generalised K\"ahler potential.
After T-dualising, this will induce a non-trivial monodromy of the dual generalised K\"ahler potential, which in terms of the dual superfields $\tilde z, \tilde w$, will not be intepretable as a generalised K\"ahler transformation. 

More exactly, we will consider potentials with monodromies of the form
\be
V(z+\bar z,w+\bar w) \rightarrow
V(z+\bar z,w+\bar w)
+ \alpha_1 (z+\bar z) 
+ \alpha_2 (w+\bar w) 
+ \gamma (z+\bar w)(w+\bar w)
\,,
\ee
with some constants $\alpha_1,\alpha_2, \gamma,\delta$.
The parameter $\gamma$ corresponds to a constant shift (large gauge transformation) of the $B$-field.
More precisely, we will phrase this monodromy as arising from some transformation acting on the real parts $y_1 \equiv z+\bar z$, $y_2 = w+\bar w$.
We will write this transformation as $(y_1,y_2) \rightarrow ( y_1^{\star}, y_2^{\star})=(y_1+n_1 ,y_2+n_2)$.
This can describe global shifts by constants $n_1,n_2$, under which we make a periodic identification, or in the non-compact case rotations by $2 \pi$ (with $n_1=n_2=0$).
We assume the potential behaves as
\be
V(y_1^{\star},y_2^{\star}) 
= V(y_1,y_2)
+\alpha_1 y_1 
+\alpha_2 y_2
+\gamma y_1 y_2 
+\delta
\,.
\label{Vshift}
\ee
We want to consider the effect of this monodromy on the dual coordinates and the dual potentials.
The Legendre transformations are:
\be
V^{(c)}(y_1,\tilde y_2) = V(y_1,y_2) + y_2 \tilde y_2\,\quad
V^{(t)}(\tilde y_1, y_2) = V(y_1,y_2) - y_1 \tilde y_1 \,.
\ee
It follows that the corresponding shifts of the dual coordinates $\tilde y_1 = V_1$ and $\tilde y_2 = - V_2$ are:
\be
\tilde y_1 \rightarrow \tilde y_1^\star = \tilde y_1 + \alpha_1 + \gamma y_2 \,,\quad
\tilde y_2 \rightarrow \tilde y_2^\star = \tilde y_2 -\alpha_2 - \gamma y_1 \,.
\ee
By carefully considering the definition of the Legendre transformed potentials, we can show (see appendix \ref{legcalc}) that:
\be
V^{(c)}(y_1^\star, \tilde y_2^\star ) 
= V^{(c)}(y_1, \tilde y_2) + \alpha_1 y_1+\delta
+ n_2 (\tilde y_2 - \alpha_2 -\gamma y_1 ) \,,
\label{Vcshift}
\ee
\be
V^{(t)}(\tilde y_1^*, y_2^*) 
= V^{(t)}(\tilde y_1, y_2) + \alpha_2 y_2 + \delta
-n_1(\tilde y_1+ \alpha_1 + \gamma y_2)\,,
\label{Vtshift}
\ee
In both \eqref{Vcshift} and \eqref{Vtshift}, the shift of the generalised K\"ahler potential is still a generalised K\"ahler transformation. 
However, if we consider the doubly dual potential, $\tV(\tilde y_1, \tilde y_2)$, we have
\be
\begin{split} 
\tV (\tilde y_1^*,\tilde y_2^*) 
& = \tV(\tilde y_1, \tilde y_2) 
- \gamma y_1(\tilde y_1, \tilde y_2) y_2(\tilde y_1, \tilde y_2) +\delta
\\ &\qquad- n_1 ( \tilde y_1 + \alpha_1 + \gamma y_2 (\tilde y_1, \tilde y_2) )
+ n_2 (\tilde y_2 - \alpha_2 - \gamma y_1(\tilde y_1, \tilde y_2)) \,.
\end{split}
\label{Vtildeshift}
\ee
When expressed in terms of the dual superfields, this is not necessarily of the form of a generalised K\"ahler transformation.
The corresponding metric and $B$-field configuration will then not describe a geometric space but rather a T-fold. 
We further see that while $\tV$ behaves less than optimally, both $V^{(c)}$ and $V^{(t)}$ transform nicely: and it is these that our doubled-generalised potential is constructed from. 
We now describe two examples illustrating these points.

\subsubsection{Torus fibration with constant H-flux and codimension-1 T-fold}

We start with the example of a torus with constant $H$-flux. It is well-known that T-duality transformations of such backgrounds can lead outside the scope of conventional geometry, producing non-geometric spaces known as T-folds.    
We consider again a setup with one chiral superfield (with real part $y_1$) and a twisted chiral superfield (with real part $y_2$) and the following potential:
\be
V(y_1,y_2) = 
\tfrac{1}{2} \left( 
(c+m y_2) y_1^2 - c y_2^2 
- \tfrac{1}{3} m y_2^3 
\right) \, , 
\ee
where $c$ and $m$ are constants. This gives metric and $B$-field
\be
ds^2 =  (c+my_2)  ( dx_1^2 + dy_1^2 +  dx_2^2 + dy_2^2 ) 
\,,\quad
B = - m y_1 d y_1 \wedge dy_2 - m y_1d x_1 \wedge dx_2 \, , 
\label{codim1}
\ee
describing the transverse geometry to a thrice-smeared NS5 brane, which is of codimension 1.
If we take $y_1, x_1,x_2$ to be periodic (with period $2\pi$) then this is a fibration of a three-torus over an interval parametrised by $y_2$. The three-torus has constant $H$-flux, $H = dB = - m dy_1 \wedge dx_1 \wedge dx_2$. 
The gauge choice for the $B$-field implies that for $y_1 \rightarrow y_1 + 2\pi$ we patch by a large gauge transformation, which corresponds to a generalised K\"ahler transformation of the potential:
\be
V(y_1+2\pi, y_2) = V(y_1,y_2) +  2\pi (c+my_2) (y_1 + \pi) \, , 
\ee 
or in complex terms
\be
\begin{split}
V(z+\bar z + 2 \pi, w+\bar w) =
V(z+\bar z, w+\bar w) 
+ 2 \pi \big(  & c\pi + c (z+\bar z) +m \pi (w+\bar w) 
\\ & \qquad + m (z w+ \bar z \bar w + z \bar w +\bar z w) 
\big)\,.
\end{split}
\ee
T-dualising the geometry \eqref{codim1} on the $x_1$ and $x_2$ directions leads to
\be
\begin{split}
\widetilde{ds}^2 & = 
f(y_2) ( d y_1^2 +  d y_2^2 ) 
+ \frac{f(y_2)}{ f(y_2)^2 + (my_1)^2  } ( d\tilde x_1^2 + d\tilde x_2^2 )\,,\\
\tilde B& =  - my_1 dy_1 \wedge dy_2 + \frac{m y_1}{  f(y_2)^2 + (my_1)^2  } d\tilde x_1 \wedge d\tilde x_2\,.
\end{split}
\ee
For $y_1 \rightarrow y_1 + 2 \pi$ this is not patched by any combination of diffeomorphisms or gauge transformations: instead the geometry is only well-defined up to a non-trivial T-duality transformation. This is associated to a so-called `exotic' $5_2^2$ brane which is non-geometric in nature (see \cite{deBoer:2012ma} for a detailed discussion).
Here we have the codimension-1 version of this exotic brane, which has been recently studied in \cite{Chaemjumrus:2019ipx} for example.

Now we consider how this T-fold behaviour manifests in the generalised K\"ahler potentials.
The Legendre dual coordinates are defined by
\be
\tilde y_1 = V_1 = f(y_2) y_1 \,,\quad
\tilde y_2 
= - V_2 = - \tfrac{1}{2} m y_1^2 + \tfrac{1}{2m} (f(y_2)^2-c^2) \,,
\label{tfoldduals}
\ee
where $f(y_2) \equiv c+my_2$.
To solve for $y_1$ and $y_2$, we write $y_1 = \tilde y_1/f$ where $f = f(y_2(\tilde y_1, \tilde y_2))$ is determined via \eqref{tfoldduals}. Explicitly:
\be
f(\tilde y_1, \tilde  y_2)^2 = m 
\left(
\tilde y_2 + \tfrac{c^2}{2m} 
\pm 
\sqrt{ \left(\tilde y_2 + \tfrac{c^2}{2m} \right)^2
+ \tilde y_1^2 }
\right)\,.
\ee
The fully T-dual potential can then be written down as
\be
\begin{split} 
\tV(\ty_1,\ty_2) 
&
= - \frac{1}{2} \frac{\ty_1^2}{f(\tilde y_1, \tilde y_2)}
+ \frac{f(\tilde y_1, \tilde y_2)-c}{m} \ty_2
\\ & \qquad 
- \frac{1}{2m^2}\left( c(f(\tilde y_1, \tilde y_2)-c)^2 - \frac{1}{3}(f(\tilde y_1, \tilde y_2)-c)^3\right)\,.
\end{split}
\ee
From \eqref{tfoldduals} we see that under $y_1 \rightarrow y_1 + 2\pi$ we have
\be
\ty_1 \rightarrow \ty_1 + 2 \pi f \,,\quad
\ty_2 \rightarrow \ty_2 - 2 m \pi ( f^{-1} \ty_1 + \pi) 
\,.
\ee
As the function $f$ is originally independent of $y_1$, it is invariant under these shifts (this can be double checked at the level of the quadratic equation solved by $f^2$ obtained from \eqref{tfoldduals} on substituting $y_1 = \ty_1 /f$ into the second equation). 
It follows that the monodromy of the dual potential is
\be
\tV \rightarrow \tV - 4 \pi \ty_1 + 2 \pi c f^{-1} \ty_1 
- 4\pi^2 f + 2 \pi^2 c \,,
\ee
which matches what one gets from the general expression \eqref{Vtildeshift}.
To emphasise the unpleasant nature of this transformation, we write it more explicitly and in terms of complex coordinates as: 
\be
\label{ugly}
\begin{split}
\tV &\rightarrow \tV
-4 \pi (\tilde w+\bar{\tilde w})
+ \frac{ 2 \pi c(\tilde w+\bar{\tilde w})}
{
\sqrt{m\left[
\tilde z+\bar{\tilde z}
+ \tfrac{c^2}{2m}\pm
\sqrt{(\tilde z+\bar{\tilde z})^2+
(\tilde w+\bar{\tilde w})^2}
\right]
}
}
\\ & \qquad
-4 \pi^2 \sqrt{m\left[
\tilde z+\bar{\tilde z}
+ \tfrac{c^2}{2m}\pm
\sqrt{(\tilde z+\bar{\tilde z})^2+
(\tilde w+\bar{\tilde w})^2}
\right]
}
+2\pi^2 c
\end{split}
\ee
In terms of the superfields $\tilde z$ and $\tilde w$, this is not a generalised K\"ahler transformation.
This reflects the non-geometric nature of the background.
We could rewrite in terms of the original superfields that
\be
\begin{split}
\tV(\tilde z, \tilde w) &\rightarrow \tV(\tilde z, \tilde w) 
- 4 \pi (c+m(w+\bar w))(z+\bar z) 
\\ & \qquad + 2\pi c(w+\bar w) 
- 4\pi^2 ( c+m(w+\bar w)) + 2 \pi^2 c
\end{split}
\ee
which would be a generalised K\"ahler transformation if we had access to the dual description involving $z$ and $w$. However, strictly speaking this requires going a doubled formalism.

Accordingly, we now construct the doubled potential $\VV=\tfrac{1}{2} (V^{(c)} + V^{(t)})$ using the two partial duals:
\be
\begin{split} 
V^{(c)} & = \frac{1}{2} \left( f(y_1,\ty_2) y_1^2 - \frac{1}{m^2} \left(
c(f(y_1,\ty_2)-c)^2 + \frac{1}{3} ( f(y_1,\ty_2)-c)^3 
\right)
\right)
\\ & \quad
+ \frac{1}{m} \ty_2 (  f(y_1,\ty_2) - c ) \,,
\\
V^{(t)}& = - \frac{1}{2} \left(
\frac{\tilde y_1^2}{c+my_2} + c y_2^2 + \tfrac{1}{3} m y_2^3
\right)\,,
\end{split}
\ee
where
\be
f(y_1,\ty_2)^2 = m^2 y_1^2 + c^2 + 2 m \ty_2^2 \,.
\ee
The monodromy transformation arises from the following shifts:
\be
y_1 \rightarrow y_1 + 2 \pi,
\quad 
\tilde y_1 \rightarrow
\tilde y_1 + 2 \pi f(y_2) \,,\quad
\tilde y_2 \rightarrow
\tilde y_2 - 2 m \pi (y_1 + \pi)\,,
\ee
in terms of which 
\be
\begin{split}
V^{(c)} & \rightarrow 
V^{(c)} + 2\pi c (y_1+\pi)\,,\\
V^{(t)} & \rightarrow V^{(t)} 
- 2\pi  (\ty_1 + \pi f(y_2)) \,.
\end{split}
\ee
It follows that $\VV$ shifts by a linear function of the coordinates which is a generalised K\"ahler transformation of the doubled-generalised potential. Once separating $\tilde{V}$ from $\VV$ These transformation turn into \eqref{ugly} upon expressing dual coordinates in terms of the physical ones.

\subsubsection{Codimension-2 T-fold}

Next, we start with the codimension-2 NS5 solution for which an $\mathcal{N}=(2,2)$ description is also available  \cite{Kiritsis:1993pb,Andreas:1998hh}.\footnote{Doubled formulations of the codimension-2 NS5 to $5_2^2$ duality in terms of $\mathcal{N}=(4,4)$ gauged linear sigma models in superspace were studied in \cite{Jensen:2011jna,Kimura:2015qze,Kimura:2018ain}.}
Let $u=y_1 + iy_2$ and let $f(u)$ be an arbitrary holomorphic function. 
Then let $V = \mathrm{Im}(f(u))$. 
The metric and $B$-field have components
\be
g_{11} = g_{22} = \mathrm{Im} f'' \,,\quad
B_{12} = - \mathrm{Re} f'' \,,
\ee
where $f''$ denotes the second derivative of $f$.
The double T-dual of this model will describe a T-fold if there is an initial monodromy giving a constant shift of the original $B$-field.
To realise this, we view $y_1$ and $y_2$ as coordinates parametrising the $\mathbb{R}^2$ transverse space of a codimension-2 brane.
We then consider functions which are not single-valued under rotations about the origin.
In particular, if $f(e^{2\pi i} u) = f(u) + \tfrac{\gamma}{2} u^2$ then under such a rotation $V \rightarrow V+ \gamma y_1 y_2$ and $B_{12} \rightarrow B_{12} - \gamma$.
A particular choice could be 
\be
f(u) = - \frac{i\gamma}{4\pi} u^2 ( \ln u - \tfrac{3}{2} )
\Rightarrow
f''(u) =  - \frac{i \gamma}{2\pi}  \ln u \,.
\ee
Writing $y_1 + iy_2 = r e^{i \theta}$, 
the geometry is thus
\be
ds^2 =  - \tfrac{\gamma}{2\pi}  \ln r \left( dr^2 + r^2 d \theta^2 + dx_1^2 + dx_2^2 \right) 
\,,\quad
B = - \tfrac{\gamma}{2\pi}  \theta r dr \wedge d\theta - \tfrac{\gamma}{2\pi}  \theta dx_1 \wedge dx_2\,.
\ee
This can be viewed as describing the geometry near a codimension-two NS5 brane.
The dual T-fold geometry is:
\be
\begin{split}
\widetilde{ds}^2 & = 
- \tfrac{\gamma}{2\pi}\ln r \left( dr^2 + r^2 d \theta^2\right) 
+ \frac{- \tfrac{\gamma}{2\pi} \ln r}{ (\tfrac{\gamma}{2\pi} \ln r)^2 + (\tfrac{\gamma}{2\pi} \theta)^2 } ( d\tilde x_1^2 + d\tilde x_2^2 )\\
\tilde B& = -\tfrac{\gamma}{2\pi} \theta r dr \wedge d\theta + \frac{\tfrac{\gamma}{2\pi} \theta}{ ( \tfrac{\gamma}{2\pi} \ln r)^2 + (\tfrac{\gamma}{2\pi}  \theta)^2 } d\tilde x_1 \wedge d\tilde x_2
\end{split}
\ee
which transforms by a non-geometric T-duality for $\theta \rightarrow \theta + 2 \pi$.

Rather than explicitly determine the dual potentials in this case, we simply note that the general discussion at the start of this subsection allows us to infer that they will inherit monodromies under 
\be
\tilde y_1 \rightarrow \tilde y_1^\star = \tilde y_1 + \gamma y_2 \,,\quad
\tilde y_2 \rightarrow \tilde y_2^\star = \tilde y_1 - \gamma y_1 \,,\quad
\ee
in terms of which $V^{(c)}$ and $V^{(t)}$ are in fact invariant
\be
V^{(c)}(y_1^\star, \tilde y_2^\star ) 
= V^{(c)}(y_1, \tilde y_2) 
\,,\quad 
V^{(t)}(\tilde y_1^*, y_2^*) 
= V^{(t)}(\tilde y_1, y_2)
\ee
but
\be
\begin{split} 
\tV (\tilde y_1^*,\tilde y_2^*) 
& = \tV(\tilde y_1, \tilde y_2) 
- \gamma y_1(\tilde y_1, \tilde y_2) y_2(\tilde y_1, \tilde y_2) \,.
\end{split}
\label{Vtildeshift22}
\ee
Accordingly while $\tV$ in this case behaves poorly under the global shift leading to the T-duality identification, the doubled-generalised K\"ahler potential built from $V^{(c)}$ and $V^{(t)}$ will be well-behaved: it is invariant. Note in this case that the transformation of the doubled coordinates $\YY^M$ leading to the monodromy is exactly that of a $B$-field shift in the original choice of polarisation with $y_1$ and $y_2$ as physical coordinates, and hence a bivector shift in the dual choice with $\tilde y_1$ and $\tilde y_2$ as physical coordinates.\footnote{This is a bivector shift acting on a geometry described by one chiral and one twisted chiral superfield, which preserves the fact that the complex structures commute. This is unlike the case of the bivector shift acting on a purely chiral or purely twisted chiral (i.e. K\"ahler) geometry, which appears in the next example, albeit not as a T-fold. It would be interesting to find explicit examples of T-folds which map from BiLP to non-BiLP geometries.}

\subsection{Semi-chiral geometry without semi-chirals} 
\label{semi}

We now consider an example where we can describe a geometry that ordinarily requires semi-chiral superfields, without introducing semi-chiral superfields. 
The starting point is the $D=2d$-dimensional K\"ahler geometry of \eqref{KahlerExample}. 
In this case, the generalised K\"ahler potential is a genuine K\"ahler potential, $V=V(y^\alpha)$, which we suppose to depend only on the real parts of $d$ chiral superfields. 
The doubled potential is then
\be
\mathbb{V}(\YY) = \tfrac{1}{2} ( V(y) + \widetilde V(\tilde y) )\,,
\ee
where $\widetilde{V}(\tilde y_a)$ is the Legendre dual of $V(y)$, and depends only on the real parts of $d$ dual twisted chiral superfields.
We recall that the constraints \eqref{cony} can be written democratically as
\be
\YY^M = \Theta^{MN} \partial_N \mathbb{V} \,,\quad \Theta^{MN} = 2 \begin{pmatrix} 0 & -\I1 \\ \I1 & 0 \end{pmatrix} \,,
\ee
implying here
\be
y^\alpha = - \widetilde{V}^\alpha,\,\quad
\tilde y_\alpha=  V_\alpha \,.
\label{yorigconstraints}
\ee
The components of the original K\"ahler metric are $
g_{\alpha \beta} = V_{\alpha \beta}$
and we have $\widetilde{V}^{\alpha \beta} = - (V^{-1})^{\alpha \beta}$.

Now we consider an $\Odd$ transformation according to the prescription of section \ref{oddtwotwo}, with $\mathbb{X}'^M = \mathcal{O}^{-1}{}^M{}_N \mathbb{X}^N$, $\mathbb{Y}'^M = \mathcal{O}^{-1}{}^M{}_N \mathbb{Y}^N$, and the bivector transformation given by
\be
\mathcal{O}^{-1}{}^M{}_N  = \begin{pmatrix} \I1 & - \beta \\
0 & \I1 
\end{pmatrix}\,,\quad
\beta \equiv \lambda \begin{pmatrix} 0 & 1 \\ -1 & 0 \end{pmatrix} \,.
\ee
Explicitly, this is (where possible we drop the indices in expresssions below):
\be
x' = x - \beta \tilde x \,,\quad
\tilde x' = \tilde x\,,
\ee
\be
y' = y - \beta \tilde y \,,\quad
\tilde y' = \tilde y\,,
\label{ybiveg}
\ee
with the doubled potential transforming as a scalar, hence
\be
\mathbb{V}'(\mathbb{Y}') = 
\tfrac{1}{2} ( V(y'+\beta \tilde y') + \widetilde{V}(\tilde y')
)\,.
\ee
The transformed constraint can be written as
\be
\mathbb{Y}'^M = \Theta'^{MN} \partial'_{N} \mathbb{V}' \,,
\ee
with
\be
\Theta'^{MN} = (\mathcal{O}^{-1} \Theta \mathcal{O}^{-T})^{MN}
= 2 \begin{pmatrix} - 2 \beta & - \I1 \\ \I1 & 0 \end{pmatrix} 
\ee
hence
\be
y'^\alpha = - \widetilde{V}^\alpha - \beta^{\alpha \beta} V_\beta \,,\quad
\tilde y'_\alpha = V_\alpha 
\label{primeconstraint}
\ee
Using \eqref{ybiveg} this can be checked to be equivalent to the original constraints \eqref{yorigconstraints}.

The initial set of superfields obey an equation of the form \eqref{overdoubledhatD}
with over-doubled complex structures $\mathbb{J}_\pm$ given by
\be
\mathbb{J}_+ = \begin{pmatrix} 0 & \I1_{2d} \\ 
- \I1_{2d} & 0 
\end{pmatrix} 
\,,\quad
\mathbb{J}_- = \begin{pmatrix} 
0 & 0 & \I1_{d} & 0 \\ 
0 & 0 & 0 & - \I1_{d} \\ 
- \I1_{d} & 0 & 0 & 0\\ 
0 &  \I1_{d} & 0 & 0\\ 
\end{pmatrix} \,.
\ee
After the $\Odd$ transformation we obtain:
\be
\mathbb{J}_+' = \begin{pmatrix} 0 & \I1_{2d } \\ 
- \I1_{2d } & 0 
\end{pmatrix} 
\,,\quad
\mathbb{J}_-' = \begin{pmatrix} 
0 & 0 & \I1_{d} & 2 \beta\\
0 & 0 & 0 & - \I1_{d } \\ 
- \I1_{d} & - 2\beta & 0 & 0\\ 
0 &  \I1_{d} & 0 & 0\\ 
\end{pmatrix} \,.
\ee
This means that while ($\tilde x'_\alpha, \tilde y'_\alpha)$ on their own obey the defining conditions of twisted chiral superfields, once we look also at ($x'^\alpha, y'^\alpha)$ we have
\be
\hat D_+ x' = + D_+ y' \,,\quad
\hat D_+ y' = - D_+ x' \,,
\label{noproblem}
\ee
\be
\hat D_-x' = + D_- y' + 2 \beta D_-\tilde x' \,,\quad
\hat D_- y' = - D_- x' - 2\beta D_- \tilde x'\,.
\label{nonstandard} 
\ee
In particular there is a mixing between the physical and dual superfields!

To make sense of this, we follow the general discussion in section \ref{JJworks} and take into account the constraints obeyed by the derivatives of the superfields.
Taking derivatives of \eqref{primeconstraint} leads as expected to
\be
D_\pm \tilde y' = (g'-b') D_\pm y'\,,\quad
D_\pm \tilde x' = \pm (g'\mp b') D_\pm x' \,,
\ee
with the correct transformations of the metric and $B$-field:
\be
g' \pm b'= (g^{-1} \pm \beta)^{-1} \,.
\ee
This agrees with \eqref{GBSCHPRIME2} on noting that the $B$-field components restricted to non-isometric directions are a total derivative. 

Using these constraints in the non-standard relationships \eqref{nonstandard} to eliminate $\tilde x_\alpha$ and $\tilde y_\alpha$, we obtain (after some algebraic manipulation):
\be
\hat D_-x' = (1+\beta g)(1-\beta g)^{-1} D_-\tilde x' \,,\quad
\hat D_- y' = - (1-\beta g)(1+\beta g)^{-1} D_- x' \,.
\label{morestandard} 
\ee
From \eqref{noproblem} and \eqref{morestandard} we can read off the spacetime complex structures $J_\pm'$ for the coordinates $(x',y')$.
They match those obtained in \eqref{JSCHPRIME2} by using a bivector $\Odd$ transformation as well as a coordinate transformation of the non-isometric directions. 
This shows that the somewhat peculiar $\Odd$ transformations of our doubled formulation correctly reproduce the usual T-duality, once the right coordinates have been identified to match with. 

This is strongly reminiscent of the work in \cite{Lindstrom:2007qf} where a potential containing semi-chiral fields was obtained as a (local) quotient of a space described solely in terms of chiral and twisted chiral superfields. This approach allows a model containing $n$ semi-chiral multiplets to be viewed as the quotient space of a model where the semi-chiral subspace was doubled and described in terms of $n$ chiral and $n$ twisted chiral superfields.

\section*{Acknowledgements}

The authors acknowledge the support of the FWO-Vlaanderen through the project G006119N, as well as that of the Vrije Universiteit Brussel through the Strategic Research Program ``High-Energy Physics''. 
CB is supported by an FWO-Vlaanderen Postdoctoral Fellowship.
DCT is supported by The Royal Society through a University Research Fellowship {\em Generalised Dualities in String Theory and Holography} URF 150185 and in part by STFC grant ST/P00055X/1.  
We would like to thank Alex Arvanitakis, Sibylle Driezen, Martin Ro\v{c}ek, David Svoboda and Rikard von Unge for discussions.  CB, AS and DCT  would like to thank the Simons Center for Geometry  and Physics for providing a stimulating environment during the conference “Generalized
Geometry and T-dualities” where some early findings of this project were presented \cite{WinNT}.

For the purpose of open access, the authors have applied a Creative Commons Attribution (CC BY) licence to any Author Accepted Manuscript version arising.

\appendix

\section{Conventions}
\label{convs}

\subsection{Worldsheet superspace conventions}
In this section we lay down two dimensional superspace conventions which are being used throughout the text. We follow the conventions used in \cite{Sevrin:2011mc} where more details are provided.

The worldsheet coordinates are denoted by $ \tau,\sigma$, using which we can define light-cone coordinates as 
\begin{equation}
\sigma^{\+}= \tau + \sigma ,\qquad \sigma ^== \tau - \sigma .\label{App1}
\end{equation}
The $\mathcal{N}=(1,1)$ fermionic coordinates are denoted by $ \theta ^+$ and $ \theta ^-$ and the
corresponding super derivatives are 
\begin{equation}
D_+ = \partial_{\theta^+} - \tfrac{\im}{2} \theta^+ \partial_{\+}
\qquad
D_- = \partial_{\theta^-} - \tfrac{\im}{2} \theta^- \partial_{=}
\end{equation}
which satisfy
\begin{equation}
D_+^2= - \frac{i}{2}\, \partial _{\+} \,,\qquad D_-^2=- \frac{i}{2}\, \partial _= \,,
\qquad \{D_+,D_-\}=0.\label{App2}
\end{equation}
The $\mathcal{N}=(1,1)$ integration measure is explicitly given by,
\be
\int d^ 2 \sigma \,d^2 \theta =\int d\tau \,d \sigma \,D_+D_- \vert
\ee
Passing from $\mathcal{N}=(1,1)$ to $ \mathcal{N}=(2,2)$ superspace requires the introduction of two more real fermionic coordinates $ \hat \theta ^+$ and $ \hat \theta ^-$
where the corresponding fermionic derivatives satisfy,
\be
\hat D_+^2= - \frac{i}{2} \,\partial _\pp \,,\qquad \hat D_-^2=- \frac{i}{2} \,\partial _= \,,
\ee
whereas all the anticommutators with the $(1,1)$ superderivative  vanish. The $\mathcal{N}=(2,2)$ integration measure is extended to,
\be
\int d^2 \sigma \,d^2 \theta \, d^2 \hat \theta =
\int d \tau\, d \sigma \,D_+D_-\, \hat D_+ \hat D_- \vert
\ee
In order to effectively work with the constraint $(2,2)$ superfields we introduce a complex basis as
\begin{equation}
\ID_\pm\equiv \hat D_\pm+i\, D_\pm \,
\qquad
\bar \ID_\pm\equiv\hat D_\pm-i\,D_\pm \,
\end{equation}
satisfying
\begin{equation}
\{\ID_+,\bar \ID_+\}= -2i\, \partial _\pp \,,
\qquad
\{\ID_-,\bar \ID_-\}= -2i\, \partial _= \,.
\end{equation}

\subsection{\texorpdfstring{$\mathcal{N}=(2,2)$}{(2,2)} superfields}
In extended superspace formulation fields have to be simplemented with constraints, here we review some of the most commonly used superfields in $N=(2,2)$ superspace occurring in the text.

\begin{enumerate}
\item Chiral field is a complex superfield $z$ satisfying the following constraints
\be
\bar\ID_\pm z=\ID_\pm\bar z=0 \,.
\ee
In terms of $N=(1,1)$ superfields $z_0$ and $\bar z_0$, this becomes,
\begin{gather}
z \,=\,z_0+i\, \hat \theta ^+D_+z_0+i\, \hat \theta ^-D_-z_0+\hat \theta ^+\hat \theta ^- D_+D_-z_0\,, \\
\bar z \,=\,\bar z_0-i\, \hat \theta ^+D_+\bar z_0-i\, \hat \theta ^-D_-\bar z_0+\hat \theta ^+\hat \theta ^- D_+D_-\bar z_0\,.
\end{gather} 
\item Twisted chiral field is a complex superfield $w$ (see e.g. \cite{Gates:1984nk}) satisfying
\be
\bar\ID_+w=\ID_-w=\ID_+\bar w=\bar\ID_-\bar w=0\,.\label{tcf}
\ee 
In terms of $N=(1,1)$ superfields $w_0$ and $\bar w_0$, this becomes,
\begin{gather}
w \,=\,w_0+i\, \hat \theta ^+D_+w_0-i\, \hat \theta ^-D_-w_0-\hat \theta ^+\hat \theta ^- D_+D_-w_0\,, \\
\bar w \,=\,\bar w_0-i\, \hat \theta ^+D_+\bar w_0+i\, \hat \theta ^-D_-\bar w_0-\hat \theta ^+\hat \theta ^- D_+D_-\bar w_0\,.
\end{gather}
\item Semi-chiral multiplet $l$, $\bar l$, $r$ and $\bar r$ \cite{Buscher:1987uw}:
\begin{equation}
\bar\ID_+l=\ID_+\bar l=\bar\ID_- r=\ID_- \bar r=0\,.\label{scf}
\end{equation}
In terms of $N=(1,1)$ superfields $r_0$ and $\bar r_0$, $\psi_-$ and $\bar\psi_-$, $l_0$ and $\bar l_0$, $\phi_+$ and $\bar\phi_+$, this becomes,
\begin{gather}
l \,=\,l_0+i\, \hat \theta ^+D_+l_0+i\, \hat \theta ^-\psi_-+\hat \theta ^+\hat \theta ^- D_+\psi_-\,, 
\\
\bar l \,=\,\bar l_0-i\, \hat \theta ^+D_+\bar l_0+i\, \hat \theta ^-\bar \psi_--\hat \theta ^+\hat \theta ^- D_+\bar \psi_-\,,
\\
r \,=\,r_0+i\, \hat \theta ^+\phi_++i\, \hat \theta ^-D_-r_0-\hat \theta ^+\hat \theta ^- D_-\phi_+\,, 
\\
\bar r \,=\,\bar r_0+i\, \hat \theta ^+\bar \phi_+-i\, \hat \theta ^-\bar D_-r_0+\hat \theta ^+\hat \theta ^- D_-\bar \phi_+\,.
\end{gather}
\end{enumerate}

\subsection{Complex structures in adapted coordinates}
\label{Jadapt}

The passage from $(1,1)$ to $(2,2)$ superspace requires a choice of a target space complex structure. However, due to the passage to adapted coordinates this dependence is usually hidden. Here we write out the dependence of the $(2,2)$ constraints on the complex structure as well as the constraints on the dual fields.     

In order to do this we rewrite the above formulas in real coordinates $X = ( x^\alpha, x^\mu, y^\alpha, y^\mu)$. The $N=(2,2)$ constraint in which the complex structures appear explicitly then reads
\be
\hat D_\pm X^I = (J_\pm)^I{}_J D_\pm X \,,
\ee
and following our superfield conventions (summarised in appendix \ref{convs}) we have in these real coordinates that
\be
J_+ = \begin{pmatrix} 0 & I_d \\ -I_d & 0 \end{pmatrix}\,,\quad
J_- = \begin{pmatrix} 0 & \sigma \\ - \sigma & 0 \end{pmatrix} \,,\quad 
{\footnotesize
\sigma \equiv \begin{pmatrix} I_{n_c} & 0  \\ 0 & - I_{n_t} \end{pmatrix} }\,.
\label{def_initialJ}
\ee
Explicitly, this means 
\be
\hat D_\pm x^\alpha = + D_\pm y^\alpha\,,\quad
\hat D_\pm y^\alpha = - D_\pm x^\alpha\,,\quad 
\hat D_\pm x^\mu = \pm D_\pm y^\mu \,,\quad
\hat D_\pm y^\mu = \mp D_\pm x^\mu \,.
\label{actuallyuseful}
\ee
For dual superfields
\be
\tilde w_\alpha = \tfrac12 ( \tilde y_\alpha + i \tilde x_\alpha) \,,\quad
\tilde z_\mu = \tfrac12 (\tilde y_\mu + i \tilde x_\mu ) \,.
\label{defdualsfields}
\ee
we have
\be
\widetilde J_+ =  - J_+^t = J_+ \,,\quad 
\widetilde J_- = + J_-^t = -J_-\,,
\label{def_dualJ}
\ee
which appear in the constraints $\hat D_\pm \widetilde X = \widetilde J_\pm D_\pm \widetilde X$ with $\widetilde X = ( \tilde x_\alpha, \tilde x_\mu, \tilde y_\alpha, \tilde y_\mu)$.
Explicitly, 
\be
\hat D_\pm \tilde x_\alpha = \pm D_\pm \tilde y_\alpha\,,\quad
\hat D_\pm \tilde y_\alpha = \mp D_\pm \tilde x_\alpha\,,\quad 
\hat D_\pm \tilde x_\mu = + D_\pm \tilde y_\mu \,,\quad
\hat D_\pm \tilde y_\mu = - D_\pm \tilde x_\mu \,.
\label{actuallyusefultilde}
\ee
In these adapted coordinates it is easy to extract expressions for the action in $\mathcal{N}=(1,1) $ superspace since 
\begin{equation}
\begin{aligned}
\mathcal{L} &= \hat{D}_+ \hat{D}_- V(X) \mid  \\
&= \partial_{IJ}V (J_+ D_+ X)^I (J_- D_- X)^J + \partial_J V  (J_- J_+)^J{}_K D_+ D_- X^K \\
&=(J_+ D_+ X)^I  V_{IJ}  (J_- D_- X)^J  - D_+X^I V_{IJ} (J_- J_+)^{J}{}_K D_-X^K \,.
\end{aligned}
\end{equation}
For the case where we have isometries such that $V = V(y^\alpha, y^\mu)$ then the Hessian matrix becomes 
\begin{equation}
\textrm{Hess} = V_{IJ} = \begin{pmatrix} 0 & 0 \\ 0 & M \end{pmatrix}\,,\quad   M =  \begin{pmatrix} V_{\alpha \beta}  & V_{\alpha \mu}   \\ V_{\mu \alpha} & V_{\mu \nu}  \end{pmatrix} \, .
\end{equation} 
Then we find 
\begin{equation}
E = G+ B= J_+^T \,\textrm{Hess}\, J_- -  \textrm{Hess}\, J_+ J_- = \begin{pmatrix} M \sigma & 0 \\ 0 & M \sigma  \end{pmatrix} \,.
\end{equation}

\section{Calculation of monodromies of Legendre transformed potentials}
\label{legcalc}
This appendix serves to verify the formula \eqref{Vcshift} for the monodromy of the Legendre transformed potential $V^{(c)}$. The calculations for $V^{(t)}$ and $\tV$ are similar.
The monodromy is inherited from the transformations $y_1^\star = y_1+n_1$ and $y_2^\star = y_2+n_2$ of the original coordinates, in terms of which $V(y_1^\star, y_2^\star)$ is given by \eqref{Vshift}. 
Given the definitions of the dual coordinates we have $\tilde y_1^\star = \tilde y_1 + \alpha_1 + \gamma y_2$, $\tilde y_2^\star = \tilde y_2 - \alpha_2 - \gamma y_1$. 
We write $V^{(c)}$ as
\be
V^{(c)}(y_1, \tilde y_2) = V(y_1, \hat y_2)  + \hat y_2 \tilde y_2 
\ee
where $\hat y_2 = \hat y_2 (y_1, \tilde y_2)$ by definition is derived from the equation $\tilde y_2 = -V_2$. 
Then, $\hat y_2^\star = \hat y_2(y_1+n_1, \tilde y_2-\alpha_2 - \gamma y_1)$ solves
\be
\tilde y_2 - \alpha_2 - \gamma y_1 = 
- V_2 (y_1+n_1 , y_2-n_2+n_2)
= - V_2(y_1, y_2-n_2) \,,
\ee
hence $\hat y_2^\star = \hat y_2 + n_2$.
It follows that 
\be
\begin{split} 
V^{(c)}(y_1^\star, \tilde y_2^\star)& = V(y_1^\star, \hat y_2^\star)  + \hat y_2^\star \tilde y_2^\star 
\\ 
& = V(y_1, \hat y_2) + \alpha_1 y_1 + \alpha_2 \hat y_2 +\gamma y_1 \hat y_2 + \delta 
+ ( \hat y_2 + n_2 ) ( \tilde y_2 - \alpha_2 - \gamma y_1 )\,.
\end{split} 
\ee
Multiplying out and cancelling leads to \eqref{Vcshift}.

\bibliography{CurrentBib.bib}

\end{document}